\documentclass{article}
\usepackage{amssymb}
\usepackage{amsfonts}
\usepackage{amsmath}

\input{tcilatex}

\begin{document}

\begin{center}
\bigskip

{\large A RENORMALIZABLE COSMODYNAMIC\ MODEL}

{\large \ }\bigskip

C. N. Ragiadakos

IEP, Tsoha 36, Athens 11521, Greece

email: crag@iep.edu.gr

\bigskip

\textbf{ABSTRACT}
\end{center}

\begin{quote}
The fermionic gyromagnetic ratio g= 2 of the Kerr-Newman spacetime cannot be
a computational "coincidence". This naturally immerges in a four dimensional
generally covariant modified Yang-Mills action, which depends on the
lorentzian complex structure of spacetime and not its metric. This metric
independence makes the model renormalizable. It is a counter example to the
general belief that "string theory is the only selfconsistent quantum model
which includes gravity". The other properties of the model are
phenomenologically very interesting too. The modified Yang-Mills action
generates a linear potential, instead of the Coulomb-like $\frac{1}{r}$
potential of the ordinary action. Therefore the Yang-Mills excitations must
be perturbatively confined. This separates the solutions of the model into
the vacuum bosonic sector of the periodic configurations, the "leptonic"
sector with fermionic solitons and their gauge field excitations, the
"hadronic" sector. Simple integrability conditions of the pure geometric
equations imply a limited number of "leptonic" and "hadronic" families. The
geometric surfaces are generally inside the SU(2,2) classical domain.
Soliton spin and gravity measure how much the surface penetrates inside the
classical domain. The i$^{0}$ point of infinity breaks the SU(2,2) symmetry
down to the Poincar\'{e} and dilation groups. A scaling breaking mechanism
is presented. Hence the pure geometric modes and asymptotically flat
solitons of the model must belong to representations of the Poincar\'{e}
group. The metrics compatible to the lorentzian complex structure are
induced by a Kaehler metric and the spacetime is a totally real lagrangian
submanifold of a Kaehler manifold. This opens up the possibility to use the
geometric quantization directly to the solitonic surfaces of the model,
considering their corresponding Kaehler symplectic manifold as their phase
space.

\newpage
\end{quote}

\renewcommand{\theequation}{\arabic{section}.\arabic{equation}}

\section{INTRODUCTION}

\setcounter{equation}{0}

The recent failure of ATLAS and CMS experiments to find minimal
supersymmetry effects and (large) higher spacetime dimensions have severe
consequences to the dominant theories of High Energy Physics. While the
discovery of the Higgs particle confirms the minimal Standard Model, its
proposed superstring extension does not seem to be compatible with the
negative results on supersymmetry. The well-known 11 dimensional superstring
model was considered as the dominant candidate for the extension of the
Standard Model to include gravity. Many proponents of the superstring model
claim that this is the \underline{unique} quantum self-consistent model,
which includes gravity. This statement is wrong. In the present review
article, I will present a renormalizable 4-dimensional generally covariant
quantum field theoretic model with first order derivatives.

The renormalizability of the model is implied by the metric independence of
its lagrangian. Recall that the linearized string action 
\begin{equation}
I_{S}=\frac{1}{2}\int d^{2}\!\xi \ \sqrt{-\gamma }\ \gamma ^{\alpha \beta }\
\partial _{\alpha }X^{\mu }\partial _{\beta }X^{\nu }\eta _{\mu \nu }
\label{i1}
\end{equation}%
has exactly the same property. It does not essentially depend on the metric $%
\gamma ^{\alpha \beta }$ of the 2-dimensional surface but on its complex
structure. It depends on its structure coordinates $(z^{0},\ z^{\widetilde{0}%
})$, because in these coordinates it takes the metric independent form%
\begin{equation}
I_{S}=\int d^{2}\!z\ \partial _{0}X^{\mu }\partial _{\widetilde{0}}X^{\nu
}\eta _{\mu \nu }  \label{i2}
\end{equation}%
All the wonderful properties of the string model are essentially based on
this characteristic feature of the string action.

The plausible question\cite{RAG1988} and exercise is \textquotedblleft what
4-dimensional action with first order derivatives depends on the complex
structure, but it does not depend on the metric of the
spacetime?\textquotedblright . The additional expectation is that such an
action may be formally renormalizable, because the regularization procedure
will not generate geometric counterterms. The term \textquotedblleft
formally\textquotedblright\ is used, because the 4-dimensional action may
have anomalies, which could destroy renormalizability, as it happens in the
string action.

The lorentzian signature of spacetime is not compatible with a \underline{%
real} tensor (complex structure) $J_{\mu }^{\;\nu }$. Therefore Flaherty
introduced a complex tensor to define the lorentzian complex structure,
which he extensively studied\cite{FLAHE1976}. It can be shown that there is
always a null tetrad $(\ell _{\mu },\,n_{\mu },\,m_{\mu },\,\overline{m}%
_{\mu })$ such that the metric tensor and the complex structure tensor take
the form

\begin{equation}
\begin{array}{l}
g_{\mu \nu }=\ell _{\mu }n_{\nu }+n_{\mu }\ell _{\nu }-m_{{}\mu }\overline{m}%
_{\nu }-\overline{m}_{\mu }m_{\nu } \\ 
\\ 
J_{\mu }^{\;\nu }=i(\ell _{\mu }n^{\nu }-n_{\mu }\ell ^{\nu }-m_{\mu }%
\overline{m}^{\nu }+\overline{m}_{\mu }m^{\nu })%
\end{array}
\label{i3}
\end{equation}%
The integrability condition of this complex structure implies the Frobenius
integrability conditions of the pairs $(\ell _{\mu },\,\,m_{\mu })$ and $%
(n_{\mu },\,\overline{m}_{\mu })$

\begin{equation}
\begin{array}{l}
(\ell ^{\mu }m^{\nu }-\ell ^{\nu }m^{\mu })(\partial _{\mu }\ell _{\nu
})=0\;\;\;\;,\;\;\;\;(\ell ^{\mu }m^{\nu }-\ell ^{\nu }m^{\mu })(\partial
_{\mu }m_{\nu })=0 \\ 
\\ 
(n^{\mu }m^{\nu }-n^{\nu }m^{\mu })(\partial _{\mu }n_{\nu
})=0\;\;\;\;,\;\;\;\;(n^{\mu }m^{\nu }-n^{\nu }m^{\mu })(\partial _{\mu
}m_{\nu })=0%
\end{array}
\label{i4}
\end{equation}%
That is, only metrics with two geodetic and shear free congruences ($\kappa
=\sigma =\lambda =\nu =0$)\cite{P-R1984} admit an integrable complex
structure.

Frobenius theorem states that there are four complex functions $%
z^{b}=(z^{\alpha },\;z^{\widetilde{\alpha }})$,\ $\alpha =0,\ 1$ , such that

\begin{equation}
dz^{\alpha }=f_{\alpha }\ \ell _{\mu }dx^{\mu }+h_{\alpha }\ m_{\mu }dx^{\mu
}\;\;\;\;,\;\;\;dz^{\widetilde{\alpha }}=f_{\widetilde{\alpha }}\ n_{\mu
}dx^{\mu }+h_{\widetilde{\alpha }}\ \overline{m}_{\mu }dx^{\mu }\;
\label{i5}
\end{equation}%
These four functions are the structure coordinates of the (integrable)
lorentzian complex structure. Notice that in the present case of lorentzian
spacetimes, the coordinates $z^{\widetilde{\alpha }}$ are not complex
conjugate of $z^{\alpha }$, because $J_{\mu }^{\;\nu }$ is no longer a real
tensor. It is exactly this complex property of $J_{\mu }^{\;\nu }$\ that
implies the pair "particle" and "antiparticle".

Using these structure coordinates, a metric independent action takes the
simple form 
\begin{equation}
\begin{array}{l}
I_{G}=\frac{1}{2}\int d^{4}\!z\det (g_{\alpha \widetilde{\alpha }})\
g^{\alpha \widetilde{\beta }}g^{\gamma \widetilde{\delta }}F_{\!j\alpha
\gamma }F_{\!j\widetilde{\beta }\widetilde{\delta }}+c.\ conj.=\int
d^{4}\!z\ F_{\!j01}F_{\!j\widetilde{0}\widetilde{1}}+c.\ c. \\ 
\\ 
F_{j_{ab}}=\partial _{a}A_{jb}-\partial _{a}A_{jb}-\gamma
\,f_{jik}A_{ia}A_{kb}%
\end{array}
\label{i6}
\end{equation}%
This transcription is possible because the metric takes the simple form $%
ds^{2}=2g_{\alpha \widetilde{\beta }}dz^{\alpha }dz^{\widetilde{\beta }}$ in
the structure coordinate system.

The covariant null tetrad form of this action\cite{RAG1990} is 
\begin{equation}
\begin{array}{l}
I_{G}=\int d^{4}\!x\ \sqrt{-g}\ \left\{ \left( \ell ^{\mu }m^{\rho
}F_{\!j\mu \rho }\right) \left( n^{\nu }\overline{m}^{\sigma }F_{\!j\nu
\sigma }\right) +\left( \ell ^{\mu }\overline{m}^{\rho }F_{\!j\mu \rho
}\right) \left( n^{\nu }m^{\sigma }F_{\!j\nu \sigma }\right) \right\} \\ 
\\ 
F_{j\mu \nu }=\partial _{\mu }A_{j\nu }-\partial _{\nu }A_{j\mu }-\gamma
\,f_{jik}A_{i\mu }A_{k\nu }%
\end{array}
\label{i7}
\end{equation}%
where $A_{j\mu }$ is an $SU(N)$ gauge field and $(\ell _{\mu },\,n_{\mu
},\,m_{\mu },\,\overline{m}_{\mu })$ is the special integrable null tetrad (%
\ref{i3}). The difference between the present action and the ordinary
Yang-Mills action becomes more clear in its following form 
\begin{equation}
I_{G}=-\frac{1}{8}\int d^{4}\!x\ \sqrt{-g}\ \left( 2g^{\mu \nu }\ g^{\rho
\sigma }-J^{\mu \nu }\ J^{\rho \sigma }-\overline{J^{\mu \nu }}\ \overline{%
J^{\rho \sigma }}\right) F_{\!j\mu \rho }F_{\!j\nu \sigma }  \label{i8}
\end{equation}%
where $g_{\mu \nu }$ is a metric derived from the null tetrad (\ref{i3}) and 
$J_{\mu }^{\;\nu }$ is the corresponding tensor of the integrable complex
structure.

In the case of the string action (\ref{i1}) we do not need additional
conditions, because any orientable 2-dimensional surface admits a complex
structure. But in the case of 4-dimensional surfaces, the integrability of
the complex structure has to be imposed through precise conditions. These
integrability conditions (\ref{i4}) may be imposed using the ordinary
procedure of Lagrange multipliers 
\begin{equation}
\begin{array}{l}
I_{C}=\int d^{4}\!x\ \{\phi _{0}(\ell ^{\mu }m^{\nu }-\ell ^{\nu }m^{\mu
})(\partial _{\mu }\ell _{\nu })+ \\ 
\\ 
\qquad +\phi _{1}(\ell ^{\mu }m^{\nu }-\ell ^{\nu }m^{\mu })(\partial _{\mu
}m_{\nu })+\phi _{\widetilde{0}}(n^{\mu }\overline{m}^{\nu }-n^{\nu }%
\overline{m}^{\mu })(\partial _{\mu }n_{\nu })+ \\ 
\\ 
\qquad +\phi _{\widetilde{1}}(n^{\mu }\overline{m}^{\nu }-n^{\nu }\overline{m%
}^{\mu })(\partial _{\mu }\overline{m}_{\nu })+c.conj.\}%
\end{array}
\label{i9}
\end{equation}%
This technique makes the complete action $I=I_{G}+I_{C}$ self-consistent and
the usual quantization procedures may be applied.

The model has been quantized using the canonical\cite{RAG1991} and the BRST%
\cite{RAG1992} procedure. After the expansion around the trivial null tetrad
(light-cone coordinates), the first order one-loop diagrams have been
computed\cite{RAG2008a} using a convenient dimensional regularization and
they were found to be finite. The model does not contain any dimensional
constant, therefore the number of the counterterms is finite. Their
independence from the metric tensor assures that there will not be geometric
term between them. Therefore no R$^{2}$ term will appear in the present
action unlike the ordinary Yang-Mills action. This assures the formal
renormalizability of the model.

The local symmetries of the action $I=I_{G}+I_{C}$ are a) the well known
local gauge transformations, b) the reparametrization symmetry as it is the
case in any generally covariant action and c) the following extended Weyl
transformation of the tetrad 
\begin{equation}
\begin{array}{l}
\ell _{\mu }^{\prime }=\Lambda \ell _{\mu }\quad ,\quad n_{\mu }^{\prime
}=Nn_{\mu }\quad ,\quad m_{\mu }^{\prime }=Mm_{\mu } \\ 
\\ 
\ell ^{\prime \mu }=\frac{1}{N}\ell ^{\mu }\quad ,\quad n^{\prime \mu }=%
\frac{1}{\Lambda }n^{\mu }\quad ,\quad m^{\prime \mu }=\frac{1}{\overline{M}}%
m^{\mu } \\ 
\\ 
\phi _{0}^{\prime }=\phi _{0}\frac{1}{\Lambda ^{2}M}\quad ,\quad \phi
_{1}^{\prime }=\phi _{1}\frac{1}{\Lambda M^{2}} \\ 
\\ 
\phi _{\widetilde{0}}^{\prime }=\phi _{\widetilde{0}}\frac{1}{N^{2}\overline{%
M}}\quad ,\quad \phi _{\widetilde{1}}^{\prime }=\phi _{\widetilde{1}}\frac{1%
}{N\overline{M}^{2}} \\ 
\\ 
g^{\prime }=g(\Lambda NM\overline{M})^{2}%
\end{array}
\label{i10}
\end{equation}%
where $\Lambda ,N$ are real functions and $M$ is a complex one. It is larger
than the ordinary Weyl (conformal) transformation. I will call this
tetrad-Weyl transformation. Then the conventional metric (\ref{i3}) takes
the form

\begin{equation}
\begin{array}{l}
g_{\mu \nu }=\Lambda N(\ell _{\mu }n_{\nu }+n_{\mu }\ell _{\nu })-M\overline{%
M}(m_{{}\mu }\overline{m}_{\nu }-\overline{m}_{\mu }m_{\nu }) \\ 
\end{array}
\label{i11}
\end{equation}

The following dimensionless geometric action term is invariant under the
tetrad-Weyl transformation.\ \ 
\begin{equation}
\begin{array}{l}
I_{g}=k\int d^{4}x\sqrt{-g}(\ell n\partial m)(\ell n\partial \overline{m})(m%
\overline{m}\partial \ell )(m\overline{m}\partial n)= \\ 
\quad =k\int d^{4}x\sqrt{-g}(\tau +\overline{\pi })(\overline{\tau }+\pi )(%
\overline{\rho }-\rho )(\overline{\mu }-\mu ) \\ 
\end{array}
\label{i12}
\end{equation}%
In the first line the null tetrad compact notation $(e_{a}e_{b}\partial
e_{c})\equiv $ $(e_{a}^{\mu }e_{b}^{\nu }-$ $e_{a}^{\nu }e_{b}^{\mu
})\partial _{\mu }e_{c\nu }$ is used and in the second line the
Newman-Penrose (NP) spin coefficient formalism is used.\ Notice that this
term is not affected (annihilated) by the integrability conditions of the
complex structure. For the sake of completeness, I will consider this term
in the derivation of the field equations, despite the fact that such a
geometric term could not be a counterterm. In any case even with this term
the model is renormalizable because $k$ is a dimensionless constant.

The mathematical formalism of the model is heavily based on the
Newman-Penrose formalism and the geometric properties of the geodetic and
shear free congruences. Therefore in any section of the present work I have
to include a short mathematical introduction in order to make it
understandable to the high energy theoretical physicists.

The most interesting direct phenomenological effects of the model are: a)
The natural emergence of the Poincar\'{e} group, instead of the BMS group in
Einstein gravity. b) The bosonic modes of the vacuum sector of the model
have 12 independent variables like the Standard Model. c) The modified
Yang-Mills action generates a linear confining potential, instead of the
Coulomb-like $\frac{1}{r}$ potential of the ordinary action. Therefore the
"colored" vacuum and solitonic excitations are perturbatively confined. d)
The solitonic sector is separated into 3+1 "lepton families" and their
confined excited colored "quark families".

It is well known that the equality of the inertial and gravitational mass is
not an accident of nature. I think that the Einstein derivation of the
equations of motion and the fermionic gyromagnetic ratio $g=2$ of the
Kerr-Newman manifold are not accidents of nature either. I think that the
present quantum field theoretic model provides the solitonic framework for
these results. Throughout this presentation I will use an elementary
particle terminology in quotes, in order to stress the analogy with current
phenomenology.

\section{FIELD\ EQUATIONS}

\setcounter{equation}{0}

The tetrad is the set of two real $e_{\mu }^{0}=\ell _{\mu }\ ,\ $\ $e_{\mu
}^{\widetilde{0}}=n_{\mu }$ and a complex vector $e_{\mu }^{1}=m_{\mu }\ ,\ $%
\ $e_{\mu }^{\widetilde{1}}=\overline{m}_{\mu }$ which are linearly
independent. Its inverse $(e_{\mu }^{a})^{-1}=e_{b}^{\nu }$ is denoted with $%
e_{0}^{\nu }=n^{\nu }\ ,\ $\ $e_{\widetilde{0}}^{\nu }=\ell ^{\nu }\ ,\ \
e_{1}^{\nu }=-\overline{m}^{\nu }\ ,\ $\ $e_{\widetilde{1}}^{\nu }=-m^{\nu }$%
. Every tetrad $e_{\mu }^{a}$ defines a metric $g_{\mu \nu }$\ relative to
which the tetrad is null. The metric (\ref{i3}) has the form $g_{\mu \nu
}=\eta _{ab}e_{\mu }^{a}e_{\nu }^{b}$ with \ 
\begin{equation}
\eta _{ab}=\eta ^{ab}=%
\begin{pmatrix}
0 & 1 & 0 & 0 \\ 
1 & 0 & 0 & 0 \\ 
0 & 0 & 0 & -1 \\ 
0 & 0 & -1 & 0%
\end{pmatrix}
\label{fe1}
\end{equation}%
Notice that the tetrad defines the metric and the precise form of $\eta
_{ab} $\ makes the tetrad null.

The NP spin coefficients\cite{P-R1984} are defined by the following
relations 
\begin{equation}
\begin{array}{l}
\nabla _{\mu }\ell _{\nu }=(\gamma +\overline{\gamma })\ell _{\mu }\ell
_{\nu }-\overline{\tau }\ell _{\mu }m_{\nu }-\tau \ell _{\mu }\overline{m}%
_{\nu }+(\varepsilon +\overline{\varepsilon })n_{\mu }\ell _{\nu }- \\ 
\qquad -\overline{\kappa }n_{\mu }m_{\nu }-\kappa n_{\mu }\overline{m}_{\nu
}-(\alpha +\overline{\beta })m_{\mu }\ell _{\nu }+\overline{\sigma }m_{\mu
}m_{\nu }+ \\ 
\qquad +\rho m_{\mu }\overline{m}_{\nu }-(\overline{\alpha }+\beta )%
\overline{m}_{\mu }\ell _{\nu }+\overline{\rho }\overline{m}_{\mu }m_{\nu
}+\sigma \overline{m}_{\mu }\overline{m}_{\nu } \\ 
\\ 
\nabla _{\mu }n_{\nu }=-(\gamma +\overline{\gamma })\ell _{\mu }n_{\nu }+\nu
\ell _{\mu }m_{\nu }+\overline{\nu }\ell _{\mu }\overline{m}_{\nu
}-(\varepsilon +\overline{\varepsilon })n_{\mu }n_{\nu }+ \\ 
\qquad +\pi n_{\mu }m_{\nu }+\overline{\pi }n_{\mu }\overline{m}_{\nu
}+(\alpha +\overline{\beta })m_{\mu }n_{\nu }-\lambda m_{\mu }m_{\nu }- \\ 
\qquad -\overline{\mu }m_{\mu }\overline{m}_{\nu }+(\overline{\alpha }+\beta
)\overline{m}_{\mu }n_{\nu }-\mu \overline{m}_{\mu }m_{\nu }-\overline{%
\lambda }\overline{m}_{\mu }\overline{m}_{\nu } \\ 
\\ 
\nabla _{\mu }m_{\nu }=\overline{\nu }\ell _{\mu }\ell _{\nu }-\tau \ell
_{\mu }n_{\nu }+(\gamma -\overline{\gamma })\ell _{\mu }m_{\nu }+\overline{%
\pi }n_{\mu }\ell _{\nu }-\kappa n_{\mu }n_{\nu }+ \\ 
\qquad +(\varepsilon -\overline{\varepsilon })n_{\mu }m_{\nu }-\overline{\mu 
}m_{\mu }\ell _{\nu }+\rho m_{\mu }n_{\nu }+(\overline{\beta }-\alpha
)m_{\mu }m_{\nu }- \\ 
\qquad -\overline{\lambda }\overline{m}_{\mu }\ell _{\nu }+\sigma \overline{m%
}_{\mu }n_{\nu }+(\overline{\alpha }-\beta )\overline{m}_{\mu }m_{\nu } \\ 
\end{array}
\label{fe2}
\end{equation}%
For their computation, it is easier to use the relations%
\begin{equation}
\begin{tabular}{|l|}
\hline
$\alpha =\frac{1}{4}[(\ell n\partial \overline{m})+(\ell \overline{m}%
\partial n)-(n\overline{m}\partial \ell )-2(m\overline{m}\partial \overline{m%
})]$ \\ \hline
$\beta =\frac{1}{4}[(\ell n\partial m)+(\ell m\partial n)-(nm\partial \ell
)-2(m\overline{m}\partial m)]$ \\ \hline
$\gamma =\frac{1}{4}[(nm\partial \overline{m})-(n\overline{m}\partial m)-(m%
\overline{m}\partial n)+2(\ell n\partial n)]$ \\ \hline
$\varepsilon =\frac{1}{4}[(\ell m\partial \overline{m})-(\ell \overline{m}%
\partial m)-(m\overline{m}\partial \ell )+2(\ell n\partial \ell )]$ \\ \hline
$\mu =-\frac{1}{2}[(m\overline{m}\partial n)+(nm\partial \overline{m})+(n%
\overline{m}\partial m)]$ \\ \hline
$\pi =\frac{1}{2}[(\ell n\partial \overline{m})-(n\overline{m}\partial \ell
)-(\ell \overline{m}\partial n)]$ \\ \hline
$\rho =\frac{1}{2}[(\ell \overline{m}\partial m)+(\ell m\partial \overline{m}%
)-(m\overline{m}\partial \ell )]$ \\ \hline
$\tau =\frac{1}{2}[(nm\partial \ell )+(\ell m\partial n)+(\ell n\partial m)]$
\\ \hline
$\kappa =(\ell m\partial \ell )\quad ,\quad \sigma =(\ell m\partial m)$ \\ 
\hline
$\nu =-(n\overline{m}\partial n)\quad ,\quad \lambda =-(n\overline{m}%
\partial \overline{m})$ \\ \hline
\end{tabular}
\label{fe3}
\end{equation}%
where the symbols $(....)$\ have been previously defined. Notice that if the
spin coefficients are defined with the last relations (\ref{fe3}), they do
not depend on a precise metric. This definition will be adopted in the
present work.

We will also use below the following commutation relations%
\begin{equation}
\begin{array}{l}
D\equiv \ell ^{\mu }\partial _{\mu }\quad ,\quad \Delta \equiv n^{\mu
}\partial _{\mu }\quad ,\quad \delta \equiv \ell ^{\mu }\partial _{\mu } \\ 
\\ 
\lbrack \Delta ,\ D]=(\gamma +\overline{\gamma })D+(\varepsilon +\overline{%
\varepsilon })\Delta -(\overline{\tau }+\pi )\delta -(\tau +\overline{\pi })%
\overline{\delta } \\ 
\\ 
\lbrack \delta ,\ D]=(\overline{\alpha }+\beta -\overline{\pi })D+\kappa
\Delta -(\overline{\rho }+\varepsilon -\overline{\varepsilon })\delta
-\sigma \overline{\delta } \\ 
\\ 
\lbrack \delta ,\ \Delta ]=-\overline{\nu }D+(\tau -\overline{\alpha }-\beta
)\Delta +(\mu -\gamma +\overline{\gamma })\delta -\overline{\lambda }%
\overline{\delta } \\ 
\\ 
\lbrack \overline{\delta },\ \delta ]=(\overline{\mu }-\mu )D+(\overline{%
\rho }-\rho )\Delta +(\alpha -\overline{\beta })\delta +(\beta -\overline{%
\alpha })\overline{\delta } \\ 
\end{array}
\label{fe4}
\end{equation}%
Besides the Weyl tensor $C_{\mu \nu \rho \sigma }$ and the Ricci $R_{\mu \nu
}$ tensor are used through their tetrad components. The Weyl tensor
components are 
\begin{equation}
\begin{array}{l}
\Psi _{0}=-C_{\mu \nu \rho \sigma }\ell ^{\mu }m^{\nu }\ell ^{\rho
}m^{\sigma }\quad ,\quad \Psi _{1}=-C_{\mu \nu \rho \sigma }\ell ^{\mu
}n^{\nu }\ell ^{\rho }m^{\sigma } \\ 
\\ 
\Psi _{3}=-C_{\mu \nu \rho \sigma }\ell ^{\mu }m^{\nu }\overline{m}^{\rho
}n^{\sigma }\quad ,\quad \Psi _{4}=-C_{\mu \nu \rho \sigma }\ell ^{\mu
}n^{\nu }\overline{m}^{\rho }n^{\sigma } \\ 
\\ 
\Psi _{5}=-C_{\mu \nu \rho \sigma }n^{\mu }\overline{m}^{\nu }n^{\rho }%
\overline{m}^{\sigma } \\ 
\end{array}
\label{fe5}
\end{equation}

All the formulas of the NP formalism may be found in the modern books of
general relativity\cite{CHAND}. This formalism is very adequate to the study
of the geodetic and shear free congruences $(\kappa =0=\sigma )$ and $(\nu
=0=\lambda )$ determined by the vectors $\ell ^{\mu }$ and $n^{\mu }$
respectively.

Variation of the action relative to the Lagrange multipliers gives back the
complex structure integrability conditions. Using the NP spin coefficients
they take the form%
\begin{equation}
\kappa =\sigma =\lambda =\nu =0  \label{fe6}
\end{equation}

Variation of the action with respect to the gauge field $A_{j\mu }$ \bigskip
gives the field equations 
\begin{equation}
\begin{array}{l}
D_{\mu }\{\sqrt{-g}[(\ell ^{\mu }m^{\nu }-\ell ^{\nu }m^{\mu })(n^{\rho }%
\overline{m}^{\sigma }F_{j\rho \sigma })+(n^{\mu }\overline{m}^{\nu }-n^{\nu
}\overline{m}^{\mu })(\ell ^{\rho }m^{\sigma }F_{j\rho \sigma })+ \\ 
\\ 
\qquad +(\ell ^{\mu }\overline{m}^{\nu }-\ell ^{\nu }\overline{m}^{\mu
})(n^{\rho }m^{\sigma }F_{j\rho \sigma })+(n^{\mu }m^{\nu }-n^{\nu }m^{\mu
})(\ell ^{\rho }\overline{m}^{\sigma }F_{j\rho \sigma })]\}=0%
\end{array}
\label{fe7}
\end{equation}%
where $D_{\mu }=\delta _{\ell j}\partial _{\mu }+\gamma f_{\ell jk}A_{k\mu }$
is the gauge symmetry covariant derivative and $\gamma $ the coupling
constant. In order to simplify the relations, I made the bracket notations $%
(e_{a}e_{b}F_{j})\equiv e_{a}^{\mu }e_{b}^{\nu }F_{j\mu \nu }\ $for the
gauge field components. Multiplying with the null tetrad, these equations
take the form%
\begin{equation}
\begin{array}{l}
\{m^{\mu }D_{\mu }+\overline{\pi }-2\overline{\alpha }\}(\ell \overline{m}%
F_{j})+\{\overline{m}^{\mu }D_{\mu }+\pi -2\alpha \}(\ell mF_{j})=0 \\ 
\\ 
\{m^{\mu }D_{\mu }+2\beta -\tau \}(n\overline{m}F_{j})+\{\overline{m}^{\mu
}D_{\mu }+2\overline{\beta }-\overline{\tau }\}(nmF_{j})=0 \\ 
\\ 
\{\ell ^{\mu }D_{\mu }+2\overline{\varepsilon }-\overline{\rho }%
\}(nmF_{j})+\{n^{\mu }D_{\mu }+\mu -2\gamma \}(\ell mF_{j})=0 \\ 
\end{array}
\label{fe8}
\end{equation}%
I put in brackets $\{...\}$ the covariant derivatives of the primary
quantities relative to the tetrad-Weyl transformations. The integrability
conditions of these field equations are satisfied identically.

Variation of the action $I=I_{G}+I_{C}+I_{g}$ with respect to the tetrad,
gives PDEs on the Lagrange multipliers. In order to preserve the relations
between the tetrad and its inverse (the covariant and contravariant forms of
the tetrad) we will use the identities 
\begin{equation}
\begin{array}{l}
\delta e_{a}^{\mu }=e_{a}^{\lambda }[-n^{\mu }\delta \ell _{\lambda }-\ell
^{\mu }\delta n_{\lambda }+\overline{m}^{\mu }\delta m_{\lambda }+m^{\mu
}\delta \overline{m}_{\lambda }] \\ 
\\ 
\delta \sqrt{-g}=\sqrt{-g}[n^{\lambda }\delta \ell _{\lambda }+\ell
^{\lambda }\delta n_{\lambda }-\overline{m}^{\lambda }\delta m_{\lambda
}-m^{\lambda }\delta \overline{m}_{\lambda }]%
\end{array}
\label{fe9}
\end{equation}%
Variation with respect to $\ell _{\lambda }$ gives the PDEs 
\begin{equation}
\begin{array}{l}
\{m^{\mu }\partial _{\mu }+3\beta -2\tau +\overline{\alpha }\}\phi _{0}+\{%
\overline{m}^{\mu }\partial _{\mu }+3\overline{\beta }-2\overline{\tau }%
+\alpha \}\overline{\phi _{0}}- \\ 
\qquad -2(nmF_{j})(n\overline{m}F_{j})+k(\tau +\overline{\pi })(\overline{%
\tau }+\pi )(\overline{\mu }-\mu )^{2}=0 \\ 
\\ 
\{\ell ^{\mu }\partial _{\mu }+3\varepsilon +\overline{\varepsilon }-\rho
\}\phi _{0}+\phi _{1}[\tau +\overline{\pi }]+(\ell nF_{j})(n\overline{m}%
F_{j})- \\ 
\qquad -k\{\overline{m}^{\mu }\partial _{\mu }+\pi +\alpha +\overline{\beta }%
-2\overline{\tau }\}[(\tau +\overline{\pi })(\overline{\tau }+\pi )(%
\overline{\mu }-\mu )]=0 \\ 
\end{array}
\label{fe10}
\end{equation}%
and the conserved current (integrability condition) 
\begin{equation}
\begin{array}{l}
\nabla _{\lambda }\{\ell ^{\lambda }[2(nmF_{j})(n\overline{m}F_{j})+\phi
_{0}(\tau -\overline{\alpha }-\beta )+\overline{\phi _{0}}(\overline{\tau }%
-\alpha -\overline{\beta })]+ \\ 
\quad +n^{\lambda }k(\tau +\overline{\pi })(\overline{\tau }+\pi )(\overline{%
\rho }-\rho )(\overline{\mu }-\mu )+ \\ 
\quad +m^{\lambda }[(\ell nF_{j})(n\overline{m}F_{j})+\phi _{0}(\varepsilon +%
\overline{\varepsilon })+\phi _{1}(\tau +\overline{\pi })+ \\ 
\quad \quad +k(\tau +\overline{\pi })(\overline{\tau }+\pi )(\overline{\mu }%
-\mu )(\overline{\tau }-\alpha -\overline{\beta })]+ \\ 
\quad +\overline{m}^{\lambda }[(\ell nF_{j})(nmF_{j})+\overline{\phi _{0}}%
(\varepsilon +\overline{\varepsilon })+\overline{\phi _{1}}(\overline{\tau }%
+\pi )- \\ 
\quad \quad -k(\tau +\overline{\pi })(\overline{\tau }+\pi )(\overline{\mu }%
-\mu )(\tau -\overline{\alpha }-\beta )]\}=0 \\ 
\end{array}
\label{fe11}
\end{equation}

Variation with respect to $n_{\lambda }$ gives the PDEs 
\begin{equation}
\begin{array}{l}
\{\overline{m}^{\mu }\partial _{\mu }-3\alpha +2\pi -\overline{\beta }\}\phi
_{\widetilde{0}}+\{m^{\mu }\partial _{\mu }-3\overline{\alpha }+2\overline{%
\pi }-\beta \}\overline{\phi _{\widetilde{0}}}- \\ 
\qquad -2(\ell mF_{j})(\ell \overline{m}F_{j})+k(\tau +\overline{\pi })(%
\overline{\tau }+\pi )(\overline{\rho }-\rho )^{2}=0 \\ 
\\ 
\{n^{\mu }\partial _{\mu }-3\gamma -\overline{\gamma }+\mu \}\phi _{%
\widetilde{0}}-\phi _{\widetilde{1}}[\overline{\tau }+\pi ]-(\ell
nF_{j})(\ell mF_{j})+ \\ 
\qquad +k\{m^{\mu }\partial _{\mu }+2\overline{\pi }-\tau -\overline{\alpha }%
-\beta \}[(\tau +\overline{\pi })(\overline{\tau }+\pi )(\overline{\rho }%
-\rho )]=0 \\ 
\end{array}
\label{fe12}
\end{equation}%
and the corresponding conserved current is%
\begin{equation}
\begin{array}{l}
\nabla _{\lambda }\{\ell ^{\lambda }k(\tau +\overline{\pi })(\overline{\tau }%
+\pi )(\overline{\rho }-\rho )(\overline{\mu }-\mu )+ \\ 
\quad +n^{\lambda }\left[ 2(\ell mF_{j})(\ell \overline{m}F_{j})+\phi _{%
\widetilde{0}}(\alpha +\overline{\beta }-\pi )+\overline{\phi _{\widetilde{0}%
}}(\overline{\alpha }+\beta -\overline{\pi })\right] - \\ 
\quad -m^{\lambda }[(\ell nF_{j})(\ell \overline{m}F_{j})+\overline{\phi _{%
\widetilde{0}}}(\gamma +\overline{\gamma })+\overline{\phi _{\widetilde{1}}}%
(\tau +\overline{\pi })- \\ 
\quad \quad -k(\tau +\overline{\pi })(\overline{\tau }+\pi )(\overline{\rho }%
-\rho )(\alpha +\overline{\beta }-\pi )]\} \\ 
\quad -\overline{m}^{\lambda }[(\ell nF_{j})(\ell mF_{j})+\phi _{\widetilde{0%
}}(\gamma +\overline{\gamma })+\phi _{\widetilde{1}}(\overline{\tau }+\pi )]+
\\ 
\quad \quad +k(\tau +\overline{\pi })(\overline{\tau }+\pi )(\overline{\rho }%
-\rho )(\overline{\alpha }+\beta -\overline{\pi })]\}=0 \\ 
\end{array}
\label{fe13}
\end{equation}%
Variation with respect to $m_{\lambda }$ gives the PDEs 
\begin{equation}
\begin{array}{l}
\{m^{\mu }\partial _{\mu }-3\overline{\alpha }+\beta +\overline{\pi }\}%
\overline{\phi _{\widetilde{1}}}+\overline{\phi _{\widetilde{0}}}[\mu -%
\overline{\mu }]-(\ell \overline{m}F_{j})(m\overline{m}F_{j})- \\ 
\quad -k\{\ell ^{\mu }\partial _{\mu }+\varepsilon -\overline{\varepsilon }-%
\overline{\rho }-2\rho \}[(\overline{\tau }+\pi )(\overline{\rho }-\rho )(%
\overline{\mu }-\mu )]=0 \\ 
\\ 
\{m^{\mu }\partial _{\mu }+3\beta -\overline{\alpha }-\tau \}\phi _{1}+\phi
_{0}[\rho -\overline{\rho }]-(n\overline{m}F_{j})(m\overline{m}F_{j})+ \\ 
\quad +k\{n^{\mu }\partial _{\mu }+2\overline{\mu }+\mu +\gamma -\overline{%
\gamma }\}[(\overline{\tau }+\pi )(\overline{\rho }-\rho )(\overline{\mu }%
-\mu )]=0 \\ 
\\ 
\{\ell ^{\mu }\partial _{\mu }+3\varepsilon -2\rho -\overline{\varepsilon }%
\}\phi _{1}+\{n^{\mu }\partial _{\mu }-3\overline{\gamma }+2\overline{\mu }%
+\gamma \}\overline{\phi _{\widetilde{1}}}- \\ 
\quad -2(\ell \overline{m}F_{j})(n\overline{m}F_{j})+k(\overline{\tau }+\pi
)^{2}(\overline{\rho }-\rho )(\overline{\mu }-\mu )=0 \\ 
\end{array}
\label{fe14}
\end{equation}%
and the corresponding conserved current is%
\begin{equation}
\begin{array}{l}
\nabla _{\lambda }\{\ell ^{\lambda }[(m\overline{m}F_{j})(n\overline{m}%
F_{j})+\phi _{0}(\overline{\rho }-\rho )+\phi _{1}(\overline{\alpha }-\beta
)+ \\ 
\quad \quad +k(\overline{\tau }+\pi )(\overline{\rho }-\rho )(\overline{\mu }%
-\mu )(\overline{\gamma }-\gamma -\overline{\mu })]+ \\ 
\quad +n^{\lambda }[(m\overline{m}F_{j})(\ell \overline{m}F_{j})+\overline{%
\phi _{\widetilde{0}}}(\overline{\mu }-\mu )+\overline{\phi _{\widetilde{1}}}%
(\overline{\alpha }-\beta )- \\ 
\quad \quad -k(\overline{\tau }+\pi )(\overline{\rho }-\rho )(\overline{\mu }%
-\mu )(\overline{\varepsilon }-\varepsilon +\rho )]- \\ 
\quad -m^{\lambda }[2(\ell \overline{m}F_{j})(n\overline{m}F_{j})+\phi _{1}(%
\overline{\varepsilon }-\varepsilon +\rho )+\overline{\phi _{\widetilde{1}}}(%
\overline{\gamma }-\gamma -\overline{\mu })]+ \\ 
\quad +\overline{m}^{\lambda }k(\tau +\overline{\pi })(\overline{\tau }+\pi
)(\overline{\rho }-\rho )(\overline{\mu }-\mu )]\}=0 \\ 
\end{array}
\label{fe15}
\end{equation}

The field equations of the model indicate the following process for their
solution. One may first solve the pure geometric equations (\ref{fe6}).
These are the vacuum configurations and possible solitonic configurations
with vanishing gauge field $F_{j\rho \sigma }$. I call this solitonic sector
"leptonic". The form of the equations indicates that for each "leptonic"
soliton, there may be solitons with non-vanishing gauge field $F_{j\rho
\sigma }$. This solitonic sector will be called "hadronic". The reason for
this name is the observation that the static potential of the gauge field
equations (\ref{fe7}) is linear\cite{RAG1999} in $r$.

We consider the trivial spherical complex structure determined by the
following (spherical) null tetrad in spherical coordinates $(t,\ r,\ \theta
,\ \varphi )$ 
\begin{equation}
\begin{array}{l}
\ell _{\mu }=\left( 1\ ,\ -1\ ,\ 0\ ,\ 0\right) \\ 
n_{\mu }=\frac{1}{2}\ \left( 1\ ,\ 1\ ,\ 0\ ,\ 0\right) \\ 
m_{\mu }=\frac{-r}{\sqrt{2}}\ \left( 0\ ,\ 0\ ,\ 1\ ,\ i\sin \theta \right)%
\end{array}
\label{fe16}
\end{equation}%
with its contravariant coordinates 
\begin{equation}
\begin{array}{l}
\ell ^{\mu }=\left( 1\ ,\ 1\ ,\ 0\ ,\ 0\right) \\ 
n^{\mu }=\frac{1}{2}\ \left( 1\ ,\ -1\ ,\ 0\ ,\ 0\right) \\ 
m^{\mu }=\frac{1}{r\sqrt{2}}\ \left( 0\ ,\ 0\ ,\ 1\ ,\ \frac{i}{\sin \theta }%
\right)%
\end{array}
\label{fe17}
\end{equation}

\bigskip If we expand the gauge field into the null tetrad%
\begin{equation}
A_{j\mu }=B_{j1}\ell _{\mu }+B_{j2}n_{\mu }+\overline{B_{j}}m_{\mu }+B_{j}%
\overline{m}_{\mu }  \label{fe18}
\end{equation}%
we find the gauge field components $B_{j1},\ B_{j2},\ B_{j}$. In the present
null tetrad, the conjugate momenta of $B_{j1},\ B_{j2}$ vanish. Therefore we
must assume $B_{j1}=0=B_{j2}$.\ Assuming the convenient gauge condition 
\begin{equation}
\overline{m}^{\nu }\partial _{\nu }\left( r\ \sin \theta \ m^{\mu }A_{j\mu
}\right) +m^{\nu }\partial _{\nu }\left( r\ \sin \theta \ m^{\mu }A_{j\mu
}\right) =0  \label{fe19}
\end{equation}%
the field equation takes the form 
\begin{equation}
\left( \frac{\partial ^{2}}{\partial t^{2}}-\frac{\partial ^{2}}{\partial
r^{2}}\right) \left( rm^{\mu }A_{j\mu }\right) =\left[ source\right]
\label{fe20}
\end{equation}%
which apparently implies a linear "gluonic" potential for the field variable 
$\left( rm^{\mu }A_{j\mu }\right) $. A more sophisticated calculation may be
done\cite{RAG2008a}.

\section{THE\ POINCARE\ GROUP}

\setcounter{equation}{0}

In the spinor formalism the tetrad takes the form 
\begin{equation}
\begin{array}{l}
\ell ^{\mu }=\frac{1}{\sqrt{2}}e_{a}^{\ \mu }\sigma _{A^{\prime }A}^{a}%
\overline{o}^{A^{\prime }}\,o^{A} \\ 
n^{\mu }=\frac{1}{\sqrt{2}}e_{a}^{\ \mu }\sigma _{A^{\prime }A}^{a}\,%
\overline{\iota }^{A^{\prime }}\iota ^{A} \\ 
m^{\mu }=\frac{1}{\sqrt{2}}e_{a}^{\ \mu }\sigma _{A^{\prime }A}^{a}\,%
\overline{\iota }^{A^{\prime }}o^{A}%
\end{array}
\label{p1}
\end{equation}%
where $e_{a}^{\ \mu }$ is any vierbein, $(o^{A},\ \iota ^{A})$ is a spinor
dyad (basis) normalized by the condition $o^{A}\iota ^{B}\epsilon _{AB}=1$
and $\sigma _{A^{\prime }A}^{0}$ is the identity and $\sigma _{A^{\prime
}A}^{i},\ i=1,2,3$ are the ordinary Pauli matrices.

In the case of a flat spacetime and the cartesian coordinates the spinorial
integrability conditions become the Kerr differential equations 
\begin{equation}
\begin{array}{l}
(\partial _{0^{\prime }0}\lambda )+\lambda (\partial _{0^{\prime }1}\lambda
)=0\ \ and\ \ (\partial _{1^{\prime }0}\lambda )+\lambda (\partial
_{1^{\prime }1}\lambda )=0 \\ 
\end{array}
\label{p2}
\end{equation}%
where $\xi ^{A}=[1,\ \lambda ]$ and the spinorial notation is used with%
\begin{equation}
\begin{array}{l}
x^{A^{\prime }A}=x^{\mu }\sigma _{\mu }^{A^{\prime }A}=%
\begin{pmatrix}
x^{0}+x^{3} & (x^{1}+ix^{2}) \\ 
(x^{1}-ix^{2}) & x^{0}-x^{3}%
\end{pmatrix}
\\ 
\\ 
x_{A^{\prime }A}=%
\begin{pmatrix}
x^{0}-x^{3} & -(x^{1}-ix^{2}) \\ 
-(x^{1}+ix^{2}) & x^{0}+x^{3}%
\end{pmatrix}
\\ 
\\ 
\partial _{A^{\prime }A}=\frac{\partial }{\partial x^{A^{\prime }A}}=\sigma
_{A^{\prime }A}^{\mu }\partial _{\mu }=%
\begin{pmatrix}
\partial _{0}+\partial _{3} & \partial _{1}-i\partial _{2} \\ 
\partial _{1}+i\partial _{2} & \partial _{0}-\partial _{3}%
\end{pmatrix}
\\ 
\end{array}
\label{p3}
\end{equation}%
In this notation primed and unprimed indices are interchanged unlike the
Penrose notation\cite{P-R1984}. Kerr's theorem states\cite{FLAHE1976} that a
general solution of these equations is any function $\lambda (x^{A^{\prime
}B})$, which satisfies a relation of the form%
\begin{equation}
K(\lambda ,\ x_{0^{\prime }0}+x_{0^{\prime }1}\lambda ,\ x_{1^{\prime
}0}+x_{1^{\prime }1}\lambda )=0  \label{p4}
\end{equation}%
where $K(\cdot ,\ \cdot ,\ \cdot )$\ is an arbitrary function.

In the general case, the integrability conditions can be formally solved
too. In every coordinate neighborhood of the spacetime, the reality
relations of the tetrad combined with (\ref{i5}) imply the following
conditions

\begin{equation}
\begin{array}{l}
dz^{0}\wedge dz^{1}\wedge d\overline{z^{0}}\wedge d\overline{z^{1}}=0 \\ 
\\ 
dz^{\widetilde{0}}\wedge dz^{\widetilde{0}}\wedge d\overline{z^{0}}\wedge d%
\overline{z^{1}}=0 \\ 
\\ 
dz^{\widetilde{0}}\wedge dz^{\widetilde{0}}\wedge d\overline{z^{\widetilde{0}%
}}\wedge d\overline{z^{\widetilde{0}}}=0%
\end{array}
\label{p5}
\end{equation}%
for the structure coordinates $z^{b}\equiv (z^{\alpha },\;z^{\widetilde{%
\alpha }})$,\ $\alpha =0,\ 1$. Hence we may conclude that there are two real
functions $\rho _{11}$ , $\rho _{22}$ and a complex one $\rho _{12}$,
defined in neighborhoods of $%
\mathbb{C}
^{4}$, such that

\begin{equation}
\rho _{11}(\overline{z^{\alpha }},z^{\alpha })=0\quad ,\quad \rho
_{12}\left( \overline{z^{\alpha }},z^{\widetilde{\alpha }}\right) =0\quad
,\quad \rho _{22}\left( \overline{z^{\widetilde{\alpha }}},z^{\widetilde{%
\alpha }}\right) =0  \label{p6}
\end{equation}%
Notice the special dependence of the defining functions on the structure
coordinates. These conditions permit us to get off the weak notion of the
lorentzian complex structure and pose it in the context of CR structure\cite%
{JACO} and real submanifolds of complex manifolds. The conditions (\ref{p6})
define totally real submanifolds\cite{BAOU} of $%
\mathbb{C}
^{4}$.

Let $z^{0}=u+iU$\ and $z^{\widetilde{0}}=v+iV$.\ Using the coordinates $%
(u,v,\zeta =z^{1})$,\ the first real condition $\rho _{11}(\overline{%
z^{\alpha }},z^{\alpha })=0$ determines $U$ and the last condition $\rho
_{22}\left( \overline{z^{\widetilde{\alpha }}},z^{\widetilde{\alpha }%
}\right) =0$ determines $V$. In the Penrose terminology the scri+ boundary
of the spacetime is $\mathfrak{J}^{\mathfrak{+}}=\{v\rightarrow \infty \
|u,\zeta =finite\}$\ and the scri- boundary is $\mathfrak{J}^{\mathfrak{-}%
}=\{u\rightarrow -\infty \ |v,\zeta =finite\}$. Asymptotic flatness of the
complex structure has to be defined with the assumptions that $\rho _{11}(%
\overline{z^{\alpha }},z^{\alpha })=0$ and $\rho _{22}\left( \overline{z^{%
\widetilde{\alpha }}},z^{\widetilde{\alpha }}\right) =0$ are compatible with
flat spacetime geodetic and shear free congruences.

Penrose noticed\cite{P-R1984} that the general solution of the Kerr theorem
takes the form%
\begin{equation}
\begin{array}{l}
\overline{X^{m}}E_{mn}X^{n}=0\quad ,\quad K(X^{m})=0 \\ 
\\ 
E_{mn}=\left( 
\begin{array}{cc}
0 & I \\ 
I & 0%
\end{array}%
\right) \\ 
\end{array}
\label{p7}
\end{equation}%
where $X^{n}$ is an element of $CP^{3}$ and $K(X^{m})$ is a homogeneous
function. Therefore, for an asymptotically flat complex structure in $%
\mathfrak{J}^{\mathfrak{+}}$ and $\mathfrak{J}^{\mathfrak{-}}$ the
conditions (\ref{p7}) may take the form%
\begin{equation}
\begin{array}{l}
\overline{X^{m1}}E_{mn}X^{n1}=0\quad ,\quad K_{1}(X^{m1})=0 \\ 
\\ 
\overline{X^{m1}}E_{mn}X^{n2}=\Omega (\overline{X^{m1}},\ X^{n2}) \\ 
\\ 
\overline{X^{m2}}E_{mn}X^{n2}=0\quad ,\quad K_{2}(X^{m2})=0 \\ 
\end{array}
\label{p8}
\end{equation}%
where $X^{mi}$ are $4\times 2$ matrices of rank-2, in order to assure the
non-degeneracy of the complex structure tensor and $\Omega (\overline{X^{m1}}%
,\ X^{n2})$\ is a homogeneous function. The structure coordinates $z^{\alpha
}$ are then two independent functions of $\frac{X^{m1}}{X^{01}}$ and $z^{%
\widetilde{\alpha }}$ are two independent functions of $\frac{X^{m2}}{X^{02}}
$.

In order to make things clear I find necessary to make very brief review of
the of the Grassmannian manifold $G_{2,2}$. We consider the set of the $%
4\times 2$ complex matrices of rank 2%
\begin{equation}
T=\left( 
\begin{array}{c}
T_{1} \\ 
T_{2}%
\end{array}%
\right)  \label{p9}
\end{equation}%
with the equivalence relation $T\sim T^{\prime }$ if there exists a $2\times
2$ invertible matrix $S$ such that%
\begin{equation}
T^{\prime }=TS  \label{p10}
\end{equation}%
This is the the $G_{2,2}$ Grassmannian manifold with coordinates%
\begin{equation}
z=T_{2}T_{1}^{-1}  \label{p11}
\end{equation}%
which completely determine the points of the set. The coordinates $T$ are
called homogeneous coordinates and the coordinates $z$ are called projective
coordinates. Under a general linear $4\times 4$ transformation%
\begin{equation}
\begin{pmatrix}
T_{1}^{\prime } \\ 
T_{2}^{\prime }%
\end{pmatrix}%
=%
\begin{pmatrix}
A_{11} & A_{12} \\ 
A_{21} & A_{22}%
\end{pmatrix}%
\left( 
\begin{array}{c}
T_{1} \\ 
T_{2}%
\end{array}%
\right)  \label{p12}
\end{equation}%
the inhomogeneous coordinates transform as%
\begin{equation}
z^{\prime }=\left( A_{21}+A_{22}\ z\right) \left( A_{11}+A_{12}\ z\right)
^{-1}  \label{p13}
\end{equation}%
which is called fractional transformation and it preserves the compact
manifold $G_{2,2}$.

The points of $G_{2,2}$ with positive definite $2\times 2$ matrix%
\begin{equation}
\begin{pmatrix}
T_{1}^{\dagger } & T_{2}^{\dagger }%
\end{pmatrix}%
\left( 
\begin{array}{cc}
I & 0 \\ 
0 & -I%
\end{array}%
\right) \left( 
\begin{array}{c}
T_{1} \\ 
T_{2}%
\end{array}%
\right) >0\quad \Longleftrightarrow \quad I-z^{\dagger }z>0  \label{p14}
\end{equation}%
is the bounded\textbf{\ }$SU(2,2)$ classical domain\cite{PYAT},\cite{XU}
because it is bounded in the general $z$-space and it is invariant under the 
$SU(2,2)$\textbf{\ }transformation%
\begin{equation}
\begin{array}{l}
\begin{pmatrix}
T_{1}^{\prime } \\ 
T_{2}^{\prime }%
\end{pmatrix}%
=%
\begin{pmatrix}
A_{11} & A_{12} \\ 
A_{21} & A_{22}%
\end{pmatrix}%
\left( 
\begin{array}{c}
T_{1} \\ 
T_{2}%
\end{array}%
\right) \\ 
\\ 
z^{\prime }=\left( A_{21}+A_{22}\ z\right) \left( A_{11}+A_{12}\ z\right)
^{-1} \\ 
\\ 
A_{11}^{\dagger }A_{11}-A_{21}^{\dagger }A_{21}=I\quad ,\quad
A_{11}^{\dagger }A_{12}-A_{21}^{\dagger }A_{22}=0 \\ 
A_{22}^{\dagger }A_{22}-A_{12}^{\dagger }A_{12}=I \\ 
\end{array}
\label{p15}
\end{equation}

The characteristic (Shilov) boundary of this domain is the $S^{1}\times
S^{3}[=U(2)]$ manifold with $z^{\dagger }z=I$.

In the homogeneous coordinates%
\begin{equation}
\begin{array}{l}
H=\left( 
\begin{array}{c}
H_{1} \\ 
H_{2}%
\end{array}%
\right) =\frac{1}{\sqrt{2}}\left( 
\begin{array}{cc}
I & -I \\ 
I & I%
\end{array}%
\right) \left( 
\begin{array}{c}
T_{1} \\ 
T_{2}%
\end{array}%
\right) \\ 
\\ 
T=\left( 
\begin{array}{c}
T_{1} \\ 
T_{2}%
\end{array}%
\right) =\frac{1}{\sqrt{2}}\left( 
\begin{array}{cc}
I & I \\ 
-I & I%
\end{array}%
\right) \left( 
\begin{array}{c}
H_{1} \\ 
H_{2}%
\end{array}%
\right)%
\end{array}
\label{p16}
\end{equation}%
because we have%
\begin{equation}
\left( 
\begin{array}{cc}
0 & I \\ 
I & 0%
\end{array}%
\right) =\frac{1}{2}\left( 
\begin{array}{cc}
I & -I \\ 
I & I%
\end{array}%
\right) \left( 
\begin{array}{cc}
I & 0 \\ 
0 & -I%
\end{array}%
\right) \left( 
\begin{array}{cc}
I & I \\ 
-I & I%
\end{array}%
\right)  \label{p17}
\end{equation}%
and the positive definite condition takes the form%
\begin{equation}
\begin{pmatrix}
H_{1}^{\dagger } & H_{2}^{\dagger }%
\end{pmatrix}%
\left( 
\begin{array}{cc}
0 & I \\ 
I & 0%
\end{array}%
\right) \left( 
\begin{array}{c}
H_{1} \\ 
H_{2}%
\end{array}%
\right) >0\quad \Longleftrightarrow \quad -i(r-r^{\dagger })=y>0
\label{p17a}
\end{equation}%
where the projective coordinate $r_{A^{\prime }B}=x_{A^{\prime
}B}+iy_{A^{\prime }B}$ is defined as $r_{A^{\prime }B}=iH_{2}H_{1}^{-1}$,
which implies $H_{2}=-irH_{1}$ and%
\begin{equation}
\begin{array}{l}
r=i(I+z)(I-z)^{-1}=i(I-z)^{-1}(I+z) \\ 
\\ 
z=(r-iI)(r+iI)^{-1}=(r+iI)^{-1}(r-iI)%
\end{array}
\label{p18}
\end{equation}%
The fractional transformations which preserve the unbounded domain are%
\begin{equation}
\begin{array}{l}
\begin{pmatrix}
H_{1}^{\prime } \\ 
H_{2}^{\prime }%
\end{pmatrix}%
=%
\begin{pmatrix}
B_{11} & B_{12} \\ 
B_{21} & B_{22}%
\end{pmatrix}%
\left( 
\begin{array}{c}
H_{1} \\ 
H_{2}%
\end{array}%
\right) \\ 
\\ 
r^{\prime }=\left( B_{22}\ r+iB_{21}\right) \left( B_{11}-iB_{12}\ r\right)
^{-1} \\ 
\\ 
B_{11}^{\dagger }B_{22}+B_{21}^{\dagger }B_{12}=I\quad ,\quad
B_{11}^{\dagger }B_{21}+B_{21}^{\dagger }B_{11}=0 \\ 
B_{22}^{\dagger }B_{12}+B_{12}^{\dagger }B_{22}=0 \\ 
\end{array}
\label{p19}
\end{equation}%
In this "upper plane" realization of the classical domain, the homogeneous
coordinates take the form $X^{mi}$\ 
\begin{equation}
X^{mi}=%
\begin{pmatrix}
\lambda ^{Ai} \\ 
-ir_{A^{\prime }B}\lambda ^{Bi}%
\end{pmatrix}
\label{p20}
\end{equation}%
and the characteristic boundary is the "real axis"%
\begin{equation}
y=0  \label{p21}
\end{equation}

The precise form \ref{p8} of the surfaces implies that the asymptotically
flat complex structures respect the $SU(2,2)$ group. This group preserves
the characteristic (Shilov) boundary $(\Omega =0)$ of\ the classical domain.
From the Penrose conformal representation of the Minkowski spacetime, we
know that it is a submanifold of this boundary. Permitting the spacetime to
have a singularity at the point $i^{0}$\ of the boundary, the $SU(2,2)$
group is broken down\cite{PYAT} to its Poincar\'{e}$\times $dilation group.
We will see below how the scaling group is expected to be spontaneously
broken.

Here I want to point out that in Einstein's gravity (with the metric been
the fundamental quantity) the asymptotically flat spacetimes belong to
representations of the BMS group\cite{P-R1984}, which does not appear in
nature.

\section{THE\ TRAJECTORY OF\ A\ "LEPTON"}

\setcounter{equation}{0}

It is clear that a pure geometric solution ($F_{j\mu \nu }=0$) should be
viewed the way Einstein-Infeld-Hoffman considered a spacetime in order to
derive the equations of motion of two bodies\cite{E-I-H1938}. In this
context the particle appears as a "concentrated gravity tube" of the
Einstein tensor $E^{\mu \nu }$ around a trajectory being the center-line of
the tube. I want to point out that solitonic configurations must be regular.
The equation of motion of two such "particles" is derived from the
self-consistency identity $\nabla _{\mu }E^{\mu \nu }\equiv 0$ on the two
coordinate neighborhoods and the definitions of center of mass and the
momenta. This great success generated the geometrodynamic ideas of Misner
and Wheeler\cite{M-W1957}.

Newman has defined a complex trajectory determined by the annihilation of
the asymptotic shear of a spacetime. Recall that a spacetime-solution of the
present model has at least two geodetic and shear free congruences. Recently
Newman and collaborators\cite{A-N-K} derived equations of motion for this
trajectory. The imaginary part of this complex trajectory has been related
to the "spin" of the "particle". They also rederived the peculiar result\cite%
{CART1968},\cite{N-W1974} that an asymptotically Kerr-Newman spacetime has
the electron gyromagnetic ratio $g=2$.

In the formula (\ref{p8}) the Kerr conditions assure the annihilation of the
shear of $\ell ^{\mu }$ and $n^{\mu }$. One can easily check that if the $%
G_{2,2}$ homogeneous coordinates $X^{mi}$ take the form\ 
\begin{equation}
X^{mi}=%
\begin{pmatrix}
\lambda ^{Ai} \\ 
-i\xi _{A^{\prime }B}^{i}(\tau _{i})\lambda ^{Bi}%
\end{pmatrix}
\label{t1}
\end{equation}%
where $\xi _{A^{\prime }B}^{i}(\tau _{i}),\ i=1,2$\ are two complex
trajectories in the Grassmannian manifold $G_{2,2}$, two Kerr functions are
derived.\ A combination of this parametrization with the Grassmannian one (%
\ref{p20}) implies the two conditions $\det [r_{A^{\prime }B}-\xi
_{A^{\prime }B}^{i}(\tau _{i})]=0$ for the two linear equations $%
[r_{A^{\prime }B}-\xi _{A^{\prime }B}^{i}(\tau _{i})]\lambda ^{Bi}=0$\ to
admit non-vanishing solutions. In this notation, the structure coordinates
are 
\begin{equation}
z^{0}=\tau _{1}\quad ,\quad z^{1}=\frac{\lambda ^{11}}{\lambda ^{01}}\quad
,\quad z^{\widetilde{0}}=\tau _{2}\quad ,\quad z^{\widetilde{1}}=-\frac{%
\lambda ^{02}}{\lambda ^{12}}  \label{t2}
\end{equation}

I do not actually know whether all the Kerr function conditions are
generated by complex trajectories, but it is clear that the present
definition of the complex trajectories is equivalent to the Newman
asymptotic definition. The $\xi _{A^{\prime }B}^{1}(\tau _{1})$\ is the
trajectory viewed from $\mathfrak{J}^{\mathfrak{+}}$ and the second $\xi
_{A^{\prime }B}^{2}(\tau _{2})$ is the trajectory viewed from $\mathfrak{J}^{%
\mathfrak{-}}$. If these two trajectories coincide, then the lorentzian
complex structure may be called "simple".

\section{STATIC\ "LEPTONIC\ SOLITONS"}

\setcounter{equation}{0}

The knowledge of the Poincar\'{e} group permit us to look for stationary
(static) axisymmetric solitonic complex structures, which will be
interpreted as particles of the model with precise mass and angular
momentum. In the case of vanishing gauge field, we may use the general
solutions (\ref{p8}) to find special solutions which respect some
symmetries. In this case the convenient coordinates are 
\begin{equation}
z^{0}=u+iU\quad ,\quad z^{1}=\zeta \quad ,\quad z^{\widetilde{0}}=v+iV\quad
,\quad z^{\widetilde{1}}=\overline{W}\ \overline{\zeta }  \label{s1}
\end{equation}%
where $u=t-r$, $v=t+r$ and $t\in R$, $r\in R$, $\zeta =e^{i\varphi }\tan 
\frac{\theta }{2}\in S^{2}$ are assumed to be the four coordinates of the
spacetime surface. Assuming the definitions 
\begin{equation}
z^{0}=i\frac{X^{21}}{X^{01}}\quad ,\quad z^{1}=\frac{X^{11}}{X^{01}}\quad
,\quad z^{\widetilde{0}}=i\frac{X^{32}}{X^{12}}\quad ,\quad z^{\widetilde{1}%
}=-\frac{X^{02}}{X^{12}}  \label{s2}
\end{equation}%
we look for massive solutions such that%
\begin{equation}
\delta X^{mi}=i\epsilon ^{0}[\mathrm{P}_{0}]_{n}^{m}X^{ni}  \label{s3}
\end{equation}%
where $\mathrm{P}_{\mu }=-\frac{1}{2}\gamma _{\mu }(1+\gamma _{5})$.\ It
implies%
\begin{equation}
\begin{array}{l}
\delta X^{0i}=0\qquad ,\qquad \delta X^{1i}=0 \\ 
\\ 
\delta X^{2i}=-i\epsilon ^{0}X^{0i}\qquad ,\qquad \delta X^{3i}=-i\epsilon
^{0}X^{1i}%
\end{array}
\label{s4}
\end{equation}%
The above definition of the structure coordinates implies%
\begin{equation}
\begin{array}{l}
\delta z^{0}=\epsilon ^{0}\qquad ,\qquad \delta z^{1}=0 \\ 
\\ 
\delta z^{\widetilde{0}}=\epsilon ^{0}\qquad ,\qquad \delta z^{\widetilde{1}%
}=0%
\end{array}
\label{s5}
\end{equation}%
and consequently%
\begin{equation}
\begin{array}{l}
\delta u=\epsilon ^{0}\qquad ,\qquad \delta U=0 \\ 
\\ 
\delta v=\epsilon ^{0}\qquad ,\qquad \delta V=0 \\ 
\\ 
\delta \zeta =0\qquad ,\qquad \delta W=0%
\end{array}
\label{s6}
\end{equation}

This procedure gives stable (time independent) solutions. We may look for
solutions, which are \textquotedblleft eigenstates" of the z-component of
the spin too. That is they are axisymmetric. In this case the homogeneous
coordinates must also satisfy the following transformations%
\begin{equation}
\delta X^{mi}=i\epsilon ^{12}[\mathrm{\Sigma }_{12}]_{n}^{m}X^{ni}
\label{s8}
\end{equation}%
where $\mathrm{\Sigma }_{\mu \nu }=\frac{1}{2}\sigma _{\mu \nu }=\frac{i}{4}%
(\gamma _{\mu }\gamma _{\nu }-\gamma _{\nu }\gamma _{\mu })$. That is we have%
\begin{equation}
\begin{array}{l}
\delta X^{0i}=-i\frac{\epsilon ^{12}}{2}X^{0i}\qquad ,\qquad \delta X^{1i}=i%
\frac{\epsilon ^{12}}{2}X^{1i} \\ 
\\ 
\delta X^{2i}=-i\frac{\epsilon ^{12}}{2}X^{2i}\qquad ,\qquad \delta X^{3i}=i%
\frac{\epsilon ^{12}}{2}X^{3i}%
\end{array}
\label{s9}
\end{equation}%
The above definition of the structure coordinates implies%
\begin{equation}
\begin{array}{l}
\delta z^{0}=0\qquad ,\qquad \delta z^{1}=i\epsilon ^{12}z^{1} \\ 
\\ 
\delta z^{\widetilde{0}}=0\qquad ,\qquad \delta z^{\widetilde{1}}=-i\epsilon
^{12}z^{\widetilde{1}}%
\end{array}
\label{s10}
\end{equation}%
and consequently%
\begin{equation}
\begin{array}{l}
\delta u=0\qquad ,\qquad \delta U=0 \\ 
\\ 
\delta v=0\qquad ,\qquad \delta V=0 \\ 
\\ 
\delta \zeta =i\epsilon ^{12}\zeta \qquad ,\qquad \delta W=0%
\end{array}
\label{s11}
\end{equation}

A general solution, which satisfies these symmetries, is given by the
relations%
\begin{equation}
\begin{array}{l}
U=U[z^{1}\overline{z^{1}}]\qquad ,\qquad V=V[z^{\widetilde{1}}\overline{z^{%
\widetilde{1}}}] \\ 
\\ 
W=W[v-u-i(V+U)]%
\end{array}
\label{s12}
\end{equation}%
Looking for an actually symmetric static quadratic polynomial Kerr function,
I found the following form 
\begin{equation}
Z^{1}Z^{2}-Z^{0}Z^{3}+2aZ^{0}Z^{1}=0  \label{s13}
\end{equation}%
This Kerr function is generated by the static trajectory 
\begin{equation}
\begin{array}{l}
\xi ^{a}(\tau )=(\tau ,\ 0,\ 0,\ ia) \\ 
\end{array}
\label{s14}
\end{equation}

The asymptotic flatness condition (\ref{p8}) implies 
\begin{equation}
\begin{array}{l}
U=-2a\frac{z^{1}\overline{z^{1}}}{1+z^{1}\overline{z^{1}}}\qquad ,\qquad V=2a%
\frac{z^{\widetilde{1}}\overline{z^{\widetilde{1}}}}{1+z^{\widetilde{1}}%
\overline{z^{\widetilde{1}}}} \\ 
\end{array}
\label{s15}
\end{equation}%
A quite general solution is found if $W\overline{W}=1$ ($V+U=0$). In this
case we have the solution%
\begin{equation}
\begin{array}{l}
U=-2a\sin ^{2}\frac{\theta }{2}\qquad ,\qquad V=2a\sin ^{2}\frac{\theta }{2}
\\ 
\\ 
W=\frac{r-ia}{r+ia}e^{-2if(r)}%
\end{array}
\label{s16}
\end{equation}%
\ One may easily compute the corresponding tetrad up to their arbitrary
factors $N_{1}$, $N_{2}$ and $N_{3}$.%
\begin{equation}
\begin{array}{l}
\ell =N_{1}[dt-dr-a\sin ^{2}\theta \ d\varphi ] \\ 
\\ 
n=N_{2}[dt+(\frac{r^{2}+a^{2}\cos 2\theta }{r^{2}+a^{2}}-2a\sin ^{2}\theta \ 
\frac{df}{dr})dr-a\sin ^{2}\theta \ d\varphi ] \\ 
\\ 
m=N_{3}[-ia\sin \theta \ (dt-dr)+(r^{2}+a^{2}\cos ^{2}\theta )d\theta
+i(r^{2}+a^{2})\sin \theta d\varphi ]%
\end{array}
\label{s17}
\end{equation}

The corresponding projective coordinates are%
\begin{equation}
\begin{array}{l}
r_{0^{\prime }0}=i\frac{X^{21}X^{12}-X^{11}X^{22}}{X^{01}X^{12}-X^{11}X^{02}}%
=\frac{z^{0}+(z^{\widetilde{0}}-2ia)z^{1}z^{\widetilde{1}}}{1+z^{1}z^{%
\widetilde{1}}} \\ 
\\ 
r_{0^{\prime }1}=i\frac{X^{01}X^{22}-X^{21}X^{02}}{X^{01}X^{12}-X^{11}X^{02}}%
=\frac{(z^{0}-z^{\widetilde{0}}+2ia)z^{\widetilde{1}}}{1+z^{1}z^{\widetilde{1%
}}} \\ 
\\ 
r_{1^{\prime }0}=i\frac{X^{31}X^{12}-X^{11}X^{32}}{X^{01}X^{12}-X^{11}X^{02}}%
=\frac{(z^{0}-z^{\widetilde{0}}+2ia)z^{1}}{1+z^{1}z^{\widetilde{1}}} \\ 
\\ 
r_{1^{\prime }1}=i\frac{X^{01}X^{32}-X^{31}X^{02}}{X^{01}X^{12}-X^{11}X^{02}}%
=\frac{z^{\widetilde{0}}+(z^{0}+2ia)z^{1}z^{\widetilde{1}}}{1+z^{1}z^{%
\widetilde{1}}}%
\end{array}
\label{s18}
\end{equation}%
If these projective coordinates become a Hermitian matrix $x_{A^{\prime }A}$%
,\ then the complex structure is compatible with the Minkowski metric.
Otherwise, it is a curved spacetime complex structure. The form (\ref{s16})
has been chosen such that for $f(r)=0$ the complex structure becomes
compatible with the Minkowski metric.

I have already showed\cite{RAG1991},\cite{RAG1999} that this tetrad takes
the following Kerr-Schild form (in the Lindquist coordinates)

\begin{equation}
\ell _{\mu }=L_{\mu }\quad ,\quad m_{\mu }=M_{\mu }\quad ,\quad n_{\mu
}=N_{\mu }+\frac{h(r)}{2(r^{2}+a^{2}\cos ^{2}\theta )}\ L_{\mu }  \label{s19}
\end{equation}%
where the null tetrad $(L_{\mu },\ N_{\mu },\ M_{\mu },\ \overline{M}_{\mu
}) $ determines the following integrable flat complex structure%
\begin{equation}
\begin{array}{l}
L_{\mu }dx^{\mu }=dt-dr-a\sin ^{2}\theta \ d\varphi \\ 
\\ 
N_{\mu }dx^{\mu }=\frac{r^{2}+a^{2}}{2(r^{2}+a^{2}\cos ^{2}\theta )}[dt+%
\frac{r^{2}+2a^{2}\cos ^{2}\theta -a^{2}}{r^{2}+a^{2}}dr-a\sin ^{2}\theta \
d\varphi ] \\ 
\\ 
M_{\mu }dx^{\mu }=\frac{-1}{\sqrt{2}(r+ia\cos \theta )}[-ia\sin \theta \
(dt-dr)+(r^{2}+a^{2}\cos ^{2}\theta )d\theta + \\ 
\qquad \qquad +i\sin \theta (r^{2}+a^{2})d\varphi ]%
\end{array}
\label{s20}
\end{equation}%
Notice that for $h(r)=-2mr+e^{2}$ the Kerr-Newman space-time is found. The
Kerr-Schild form has the following NP spin coefficients%
\begin{equation}
\begin{tabular}{|l|}
\hline
$\alpha =\frac{ia(1+\sin ^{2}\theta )-r\cos \theta }{2\sqrt{2}\sin \theta \
(r-ia\cos \theta )^{2}}\quad ,\quad \beta =\frac{\cos \theta }{2\sqrt{2}\sin
\theta \ (r+ia\cos \theta )}$ \\ \hline
$\gamma =-\frac{a^{2}+iar\cos \theta +h}{2\rho ^{2}\ (r-ia\cos \theta )}+%
\frac{h^{\prime }}{4\rho ^{2}}\quad ,\quad \varepsilon =0$ \\ \hline
$\mu =-\frac{r^{2}+a^{2}+h}{2\rho ^{2}\ (r-ia\cos \theta )}\quad ,\quad \pi =%
\frac{ia\sin \theta }{\sqrt{2}(r-ia\cos \theta )^{2}}$ \\ \hline
$\rho =-\frac{1}{r-ia\cos \theta }\quad ,\quad \tau =-\frac{ia\sin \theta }{%
\sqrt{2}\rho ^{2}}$ \\ \hline
$\kappa =0\quad ,\quad \sigma =0\quad ,\quad \nu =0\quad ,\quad \lambda =0$
\\ \hline
\end{tabular}
\label{s21}
\end{equation}

The soliton form factor $f(r)$\ is expected to be fixed by Quantum Theory,
but I have not yet found the precise procedure. The massive configuration (%
\ref{s19}) with spin $S_{z}=ma=\frac{h}{2}$ has $g=2$ gyromagnetic ratio.
Notice that the fact that the lorentzian complex structure is a complex
tensor implies that the complex conjugate structure defines an independent
lorentzian complex structure with the same mass, which I will call
"antiparticle". This natural differentiation between "particles" and
"antiparticles" makes the complex structure more convenient than the metric
to describe elementary particles. On the other hand we know that the common
point $i^{0}$ of the $\mathfrak{J}^{\mathfrak{+}}$\ and $\mathfrak{J}^{%
\mathfrak{-}}$\ at infinity is a singular point\cite{P-R1984}. This implies
that these configurations will belong into two representations of the Poincar%
\'{e} group.

\section{THE\ "LEPTONIC\ FAMILIES"}

\setcounter{equation}{0}

The integrability conditions of the complex structure can be formulated in
the spinor formalism. They imply that both spinors $o^{A}$ and $\iota ^{A}$
of the dyad satisfy the same PDE

\begin{equation}
\xi ^{A}\xi ^{B}\nabla _{A^{\prime }A}\ \xi _{B}=0  \label{f1}
\end{equation}%
where $\nabla _{A^{\prime }A}$ is the covariant derivative connected to the
vierbein $e_{a}^{\ \mu }$. The integrability condition of these relations is%
\cite{P-R1984}

\begin{equation}
\Psi _{ABCD}\xi ^{A}\xi ^{B}\xi ^{C}\xi ^{D}=0  \label{f2}
\end{equation}%
In a curved spacetime, which admits a complex structure, the geodetic and
shear free congruences are determined by the solutions of the above 4$^{th}$
degree polynomial, which satisfy the integrability conditions (\ref{f1}). Or
vice-versa, the geodetic and shear free congruences must coincide with the
principal directions of the Weyl spinor $\Psi _{ABCD}$, because in this
tetrad we have $\Psi _{0}=0=\Psi _{4}$. Hence a spacetime with non-vanishing
Weyl tensor may admit a limited number of complex structures and their
classification coincides with the well-known Petrov classification
restricted to spacetimes which admit two geodetic and shear free
congruences. The number of $\xi ^{A}(x)$ sheets, that a regular manifold
admits, is a topological invariant. Taking into account that we need two
sheets $(o^{A},\ \iota ^{A})$ to determine a complex structure, we have the
following four cases\ \ 
\begin{equation}
\begin{array}{l}
Case\ I:\Psi _{1}\neq 0\ ,\ \Psi _{2}\neq 0\ ,\ \Psi _{3}\neq 0 \\ 
\\ 
Case\ II:\Psi _{1}\neq 0\ ,\ \Psi _{2}\neq 0\ ,\ \Psi _{3}=0 \\ 
\\ 
Case\ III:\Psi _{1}\neq 0\ ,\ \Psi _{2}=0\ ,\ \Psi _{3}=0 \\ 
\\ 
Case\ D:\Psi _{1}=0\ ,\ \Psi _{2}\neq 0\ ,\ \Psi _{3}=0 \\ 
\end{array}
\label{f3}
\end{equation}%
The type N spacetimes do not admit a complex structure.

The four sheets on the regular spacetimes are expected to generate branch
"surfaces" where the geodetic congruences will pass from the one to the
other. The well known ring singularity of the Kerr-like complex structures
are such branch "surfaces".

\section{VACUUM\ AND\ SOLITON\ SECTORS}

\setcounter{equation}{0}

Recall that the two dimensional $\phi ^{4}$-model\cite{FELS1981} has two
vacua with $\phi =\pm \frac{\mu }{\sqrt{\lambda }}$. It is well known that
the vacuum configurations are periodic, while the soliton configurations are
not periodic. This characteristic difference will be used in the present
model. The kink configuration and its excitations satisfy the boundary
conditions $\phi _{kink}(\pm \infty ,t)=\pm \frac{\mu }{\sqrt{\lambda }}$
and the antikink configuration the opposite ones.

The vacuum sector of the model are spacetimes which admit two geodetic and
shear free congruences (GSFC) which become periodic after the identification
of $\mathfrak{J}^{\mathfrak{+}}$ and $\mathfrak{J}^{\mathfrak{-}}$.
Minkowski spacetime and its smooth deformations satisfy this periodicity
criterion. Non periodic spacetimes constitute the solitonic sectors of the
model, which we will call "leptons".

We must be careful to apply the periodicity criterion to the lorentzian
complex structure (the GSFC) and not a precise metric of the spacetime. It
is well known\cite{P-R1984} that a mass term implies non-periodicity of the
metric. But it does not mean that $\ell ^{\mu },n^{\mu }$ of the two GSFCs
are not periodic up to a tetrad-Weyl and diffeomorphic transformation.
Typical examples are the massive spherically symmetric metrics which are not
periodic. But these spacetimes are equivalent to Minkowski spacetime up a
tetrad-Weyl transformation. Hence their GSFCs are periodic.

In order to make things explicit the Kerr-Newman integrable null tetrad will
be used as an example. Around $\mathfrak{I}^{\mathfrak{+}}$ the coordinates $%
(u,\ w=\frac{1}{r},\ \theta ,\ \varphi )$ are used, where the integrable
tetrad takes the form

\begin{equation}
\begin{array}{l}
\ell =du-a\sin ^{2}\theta \ d\varphi \\ 
\\ 
n=\frac{1-2mw+e^{2}w^{2}+a^{2}w^{2}}{2w^{2}(1+a^{2}w^{2}\cos ^{2}\theta )}%
[w^{2}\ du-\frac{2(1+a^{2}w^{2}\cos ^{2}\theta )}{1-2mw+e^{2}w^{2}+a^{2}w^{2}%
}\ dw-aw^{2}\sin ^{2}\theta \ d\varphi ] \\ 
\\ 
m=\frac{1}{\sqrt{2}w(1+iaw\cos \theta )}[iaw^{2}\sin \theta \
du-(1+a^{2}w^{2}\cos ^{2}\theta )\ d\theta - \\ 
\qquad -i\sin \theta (1+aw^{2})\ d\varphi ]%
\end{array}
\label{v1}
\end{equation}%
The physical space is for $w>0$\ and the integrable tetrad is regular on $%
\mathfrak{I}^{\mathfrak{+}}$ up to a factor, which does not affect the
congruence, and\ it can be regularly extended to $w<0$. Around $\mathfrak{I}%
^{\mathfrak{-}}$ the coordinates $(v,\ w^{\prime },\ \theta ^{\prime },\
\varphi ^{\prime })$ are used with

\begin{equation}
\begin{array}{l}
dv=du+\frac{2(r^{2}+a^{2})}{r^{2}-2mr+e^{2}+a^{2}}\ dr \\ 
\\ 
dw^{\prime }=-dw\quad ,\quad d\theta ^{\prime }=d\theta \\ 
\\ 
d\varphi ^{\prime }=d\varphi +\frac{2a}{r^{2}-2mr+e^{2}+a^{2}}\ dr%
\end{array}
\label{v2}
\end{equation}%
and the integrable tetrad takes the form

\begin{equation}
\begin{array}{l}
\ell =\frac{1}{w^{\prime 2}}[w^{\prime 2}\ dv-\frac{2(1+a^{2}w^{\prime
2}\cos ^{2}\theta )}{1+2mw^{\prime }+e^{2}w^{\prime 2}+a^{2}w^{\prime 2}}\
dw^{\prime }-aw^{\prime 2}\sin ^{2}\theta ^{\prime }\ d\varphi ^{\prime }]
\\ 
\\ 
n=\frac{1+2mw^{\prime }+e^{2}w^{\prime 2}+a^{2}w^{\prime 2}}{%
2(1+a^{2}w^{\prime 2}\cos ^{2}\theta ^{\prime })}[dv-a\sin ^{2}\theta
^{\prime }\ d\varphi ^{\prime }] \\ 
\\ 
m=\frac{-1}{\sqrt{2}w^{\prime }(1-iaw^{\prime }\cos \theta ^{\prime })}%
[iaw^{\prime 2}\sin \theta \ dv-(1+a^{2}w^{\prime 2}\cos ^{2}\theta ^{\prime
})\ d\theta ^{\prime }- \\ 
\qquad -i\sin \theta ^{\prime }(1+aw^{\prime 2})\ d\varphi ^{\prime }]%
\end{array}
\label{v3}
\end{equation}%
The physical space is for $w<0$\ and the integrable tetrad is regular on $%
\mathfrak{I}^{\mathfrak{-}}$ up to a factor, which does not affect the
congruence, and\ it can be regularly extended to $w>0$. If the mass term
vanishes the two regions $\mathfrak{I}^{\mathfrak{+}}$ and $\mathfrak{I}^{%
\mathfrak{-}}$ can be identified\ and the $\ell ^{\mu }$ and $n^{\mu }$
congruences are interchanged, when $\mathfrak{I}^{\mathfrak{+}}$ $(\equiv 
\mathfrak{I}^{\mathfrak{-}})$ is crossed. When $m\neq 0$\ these two regions
cannot be identified and the complex structure cannot be extended across $%
\mathfrak{I}^{\mathfrak{+}}$ and $\mathfrak{I}^{\mathfrak{-}}$.

In the present model the two real $\ell ^{\mu },n^{\mu }$ and the complex $%
m^{\mu }$ vector fields of the tetrad characterize the lorentzian complex
structure. We already know that the excitation modes must belong into
unitary representations of the Poincar\'{e} group. I have not yet found a
formal definition of the excitation modes, but they must not have more than
12 independent variables. Notice that the Standard Model bosonic modes
(Higg's particle, $\gamma $ , $Z$ , $W$) are exactly 12.

\section{THE\ DILATION\ BREAKING MECHANISM}

\setcounter{equation}{0}

It has already pointed out\cite{N-N2005},\cite{A-N2009} that the explicit
conditions $\rho _{11}(\overline{z^{\alpha }},z^{\alpha })=0\ ,\ \rho
_{22}\left( \overline{z^{\widetilde{\alpha }}},z^{\widetilde{\alpha }%
}\right) =0$\ and the corresponding holomorphic transformations $z^{\prime
\alpha }=f^{\alpha }(z^{\alpha })$ and $z^{\prime \widetilde{\alpha }}=f^{%
\widetilde{\alpha }}(z^{\widetilde{\alpha }})$ which preserve the lorentzian
complex structure, are exactly those of the 3-dimensional CR structures\cite%
{JACO1990}. Therefore we may use the Moser procedure for the classification%
\cite{RAG2011} of the lorentzian complex structures. For each hypersurface
type CR nondegenerate structure we consider the following Moser expansions \ 
\begin{equation}
\begin{array}{l}
U=z^{1}\overline{z^{1}}+\tsum\limits_{k\geq 2,j\geq 2}N_{jk}(u)(z^{1})^{j}(%
\overline{z^{1}})^{k} \\ 
N_{22}=N_{32}=N_{33}=0 \\ 
V=z^{\widetilde{1}}\overline{z^{\widetilde{1}}}+\tsum\limits_{k\geq 2,j\geq
2}\widetilde{N}_{jk}(v)(z^{\widetilde{1}})^{j}(\overline{z^{\widetilde{1}}}%
)^{k} \\ 
\widetilde{N}_{22}=\widetilde{N}_{32}=\widetilde{N}_{33}=0%
\end{array}
\label{d1}
\end{equation}%
where $z^{0}=u+iU\ ,\ z^{\widetilde{0}}=v+iV$\ and the functions $N_{jk}(u)\
,\ \widetilde{N}_{jk}(v)$ characterize the lorentzian complex structure. By
their construction these functions belong into representations of the
isotropy subgroup of $SU(1,2)$ symmetry group of the hyperquadric. Notice
that the corresponding Moser chains are determined by $n^{\alpha }\frac{%
\partial }{\partial z^{\alpha }}$ and $\ell ^{\widetilde{\alpha }}\frac{%
\partial }{\partial z\widetilde{^{\alpha }}}$.

These Moser expansions hide a dilation symmetry breaking, because the
coefficients of the first term $z^{1}\overline{z^{1}}$ ($z^{\widetilde{1}}%
\overline{z^{\widetilde{1}}}$) in the $U$ ($V$) expansion is assumed to be
non-vanishing. This is implied by the nondegeneracy condition on CR
structure. It is known that these coefficients are relative invariants of
the corresponding CR structures. On the other hand these coefficients have
the "length" dimension. Therefore, expanding a nondegenerate CR structure we
have to fix these dimensional parameters, which implies scaling symmetry
breaking.

In order to make things clear I will approach the same problem using the
differential forms of the tetrad. \ 
\begin{equation}
\begin{array}{l}
d\ell =(\varepsilon +\overline{\varepsilon })n\wedge \ell +(\overline{\tau }%
-\alpha -\overline{\beta })m\wedge \ell +(\tau -\overline{\alpha }-\beta )%
\overline{m}\wedge \ell + \\ 
\qquad +(\rho -\overline{\rho })m\wedge \overline{m}-\overline{\kappa }%
n\wedge m-\kappa n\wedge \overline{m} \\ 
\\ 
dn=-(\gamma +\overline{\gamma })\ell \wedge n+(\alpha +\overline{\beta }-\pi
)m\wedge n+(\overline{\alpha }+\beta -\overline{\pi })\overline{m}\wedge n+
\\ 
\qquad +(\mu -\overline{\mu })m\wedge \overline{m}+\nu \ell \wedge m+%
\overline{\nu }\ell \wedge \overline{m} \\ 
\\ 
dm=(\gamma -\overline{\gamma }+\overline{\mu })\ell \wedge m+(\varepsilon -%
\overline{\varepsilon }-\rho )n\wedge m+(\overline{\alpha }-\beta )\overline{%
m}\wedge m- \\ 
\qquad -(\tau +\overline{\pi })\ell \wedge n+\overline{\lambda }\ell \wedge 
\overline{m}-\sigma n\wedge \overline{m} \\ 
\end{array}
\label{d2}
\end{equation}%
It is integrable if $\kappa =\sigma =\lambda =\nu =0$.\ 

Under tetrad Weyl transformations the Newman-Penrose spin coefficients\cite%
{CHAND} transform as follows%
\begin{equation}
\begin{tabular}{l}
$\alpha ^{\prime }=\frac{1}{M}\alpha +\frac{M\ \overline{M}-\Lambda N}{%
4M\Lambda N}(\overline{\tau }+\pi )+\frac{1}{4M}\overline{\delta }\ln \frac{%
\Lambda }{N\overline{M}^{2}}$ \\ 
$\beta ^{\prime }=\frac{1}{\overline{M}}\beta +\frac{M\ \overline{M}-\Lambda
N}{4\overline{M}\Lambda N}(\tau +\overline{\pi })+\frac{1}{4\overline{M}}%
\delta \ln \frac{\Lambda M^{2}}{N}$ \\ 
$\gamma ^{\prime }=\frac{1}{\Lambda }\gamma +\frac{M\ \overline{M}-\Lambda N%
}{4M\ \overline{M}\Lambda }(\overline{\mu }-\mu )+\frac{1}{4\Lambda }\Delta
\ln \frac{M}{N^{2}\overline{M}}$ \\ 
$\varepsilon ^{\prime }=\frac{1}{N}\varepsilon +\frac{M\ \overline{M}%
-\Lambda N}{4M\ \overline{M}N}(\overline{\rho }-\rho )+\frac{1}{4N}D\ln 
\frac{M\Lambda ^{2}}{\overline{M}}$ \\ 
$\mu ^{\prime }=\frac{1}{2\Lambda }(\mu +\overline{\mu })+\frac{N}{2M\ 
\overline{M}}(\mu -\overline{\mu })+\frac{1}{2\Lambda }\Delta \ln (M\ 
\overline{M})$ \\ 
$\rho ^{\prime }=\frac{1}{2N}(\rho +\overline{\rho })+\frac{\Lambda }{2M\ 
\overline{M}}(\rho -\overline{\rho })-\frac{1}{2N}D\ln (M\ \overline{M})$ \\ 
$\pi ^{\prime }=\frac{\overline{M}}{2\Lambda N}(\pi +\overline{\tau })+\frac{%
1}{2M}(\pi -\overline{\tau })+\frac{1}{2M}\overline{\delta }\ln (\Lambda N)$
\\ 
$\tau ^{\prime }=\frac{M}{2\Lambda N}(\tau +\overline{\pi })+\frac{1}{2%
\overline{M}}(\tau -\overline{\pi })-\frac{1}{2\overline{M}}\delta \ln
(\Lambda N)$ \\ 
$\kappa ^{\prime }=\frac{\Lambda }{N\overline{M}}\kappa \quad ,\quad \sigma
^{\prime }=\frac{M}{N\overline{M}}\sigma \quad ,\quad \nu ^{\prime }=\frac{N%
}{\Lambda M}\nu \quad ,\quad \lambda ^{\prime }=\frac{\overline{M}}{\Lambda M%
}\lambda $%
\end{tabular}
\label{d3}
\end{equation}%
We see that $(\rho -\overline{\rho })\ ,(\mu -\overline{\mu })\ ,\ (\tau +%
\overline{\pi })$ undergo the multiplicative transformations 
\begin{equation}
\begin{tabular}{l}
$\rho ^{\prime }-\overline{\rho ^{\prime }}=\frac{\Lambda }{M\overline{M}}%
(\rho -\overline{\rho })$ \\ 
$\mu ^{\prime }-\overline{\mu ^{\prime }}=\frac{N}{M\overline{M}}(\mu -%
\overline{\mu })$ \\ 
$\tau ^{\prime }+\overline{\pi ^{\prime }}=\frac{M}{\Lambda N}(\tau +%
\overline{\pi })$%
\end{tabular}
\label{d4}
\end{equation}%
It implies that the vanishing or not of $(\rho -\overline{\rho })\ ,(\mu -%
\overline{\mu })\ ,\ (\tau +\overline{\pi })$ are relative invariants of the
lorentzian complex structure. If these quantities vanish, the complex
structure is kaehlerian, and the vectors of the null tetrad are hypersurface
orthogonal. That is the lorentzian complex structure is trivial and
apparently compatible with the Minkowski metric.

Taking into account the [length] dimensionality of $(\rho -\overline{\rho }%
)\ ,(\mu -\overline{\mu })\ ,\ (\tau +\overline{\pi })$ we may conclude that
if the vacuum lorentzian complex structure has at least one of them which
does not vanish, the scaling symmetry is broken.

As I have already pointed out, the tetrad-Weyl transformation should be
considered the natural extension in four dimensions of the powerful two
dimensional conformal transformation. Therefore it would be interesting to
find an analogous formulation in four dimensions. Under tetrad-Weyl
transformation (\ref{d3}) a primary field $\phi (x)$ of weight $w=(w_{1},\
w_{2},\ w_{3},\ w_{4})$ transforms as \ 
\begin{equation}
\begin{array}{l}
\phi ^{\prime }=\Lambda ^{w_{1}}N^{w_{2}}M^{w_{3}}\overline{M}^{w_{4}}\phi
\\ 
\end{array}
\label{d5}
\end{equation}%
The covariant derivative is defined with the aid of two real $Z_{1\mu },\
Z_{2\mu }$ and a complex vector field $Z_{\mu }$\ such that \ 
\begin{equation}
\begin{array}{l}
\widehat{D}_{\mu }\phi =(\partial _{\mu }-w_{1}Z_{1\mu }-w_{2}Z_{2\mu
}-w_{3}Z_{\mu }-w_{4}Z_{\mu })\phi \\ 
\end{array}
\label{d6}
\end{equation}%
and their transformations are \ 
\begin{equation}
\begin{array}{l}
Z_{1\mu }^{\prime }=Z_{1\mu }-\partial _{\mu }\Lambda \quad ,\quad Z_{2\mu
}^{\prime }=Z_{2\mu }-\partial _{\mu }N\quad ,\quad Z_{\mu }^{\prime
}=Z_{\mu }-\partial _{\mu }M \\ 
\end{array}
\label{d7}
\end{equation}%
The geometric combinations of the spin coefficients which satisfy these
gauge transformations are \ 
\begin{equation}
\begin{array}{l}
Z_{1\mu }=(\theta _{1}+\mu +\overline{\mu })\ell _{\mu }+(\varepsilon +%
\overline{\varepsilon })n_{\mu }-(\alpha +\overline{\beta }-\overline{\tau }%
)m_{\mu }-(\overline{\alpha }+\beta -\tau )\overline{m}_{\mu } \\ 
\\ 
Z_{2\mu }=-(\gamma +\overline{\gamma })\ell _{\mu }+(\theta _{2}-\rho -%
\overline{\rho })n_{\mu }-(\pi -\alpha -\overline{\beta })m_{\mu }-(%
\overline{\pi }-\overline{\alpha }-\beta )\overline{m}_{\mu } \\ 
\\ 
Z_{\mu }=(\gamma -\overline{\gamma }+\overline{\mu })\ell _{\mu
}+(\varepsilon -\overline{\varepsilon }-\rho )n_{\mu }-(\theta +\pi -%
\overline{\tau })m_{\mu }-(\beta -\overline{\alpha })\overline{m}_{\mu } \\ 
\end{array}
\label{d8}
\end{equation}%
where the additional geometric quantities are \ 
\begin{equation}
\begin{array}{l}
\theta _{1}=n^{\mu }\partial _{\mu }\ln \frac{\rho -\overline{\rho }}{2i}%
\quad ,\quad \theta _{2}=\ell ^{\mu }\partial _{\mu }\ln \frac{\mu -%
\overline{\mu }}{2i}\quad ,\quad \theta =\overline{m}^{\mu }\partial _{\mu
}\ln (\tau +\overline{\pi }) \\ 
\end{array}
\label{d9}
\end{equation}%
Notice their logarithmic dependence and that the field equations do not
contain these quantities.

\section{THE KAEHLER\ AMBIENT\ MANIFOLD}

\setcounter{equation}{0}

The four real conditions (\ref{p6}) imply that the spacetime, which admits
an integrable lorentzian complex structure, is a CR manifold with
codimension four. Following the ordinary procedure\cite{BAOU} we can find
the corresponding four real forms. It is convenient to use the notation $%
\partial f=\frac{\partial f}{\partial z^{\alpha }}dz^{\alpha }$\ and $%
\widetilde{\partial }f=\frac{\partial f}{\partial z^{\widetilde{\alpha }}}%
dz^{\widetilde{\alpha }}$. Assuming a restriction to the submanifold we find
\ 
\begin{equation}
\begin{array}{l}
\ell =2i\partial \rho _{11}|_{S}=i(\partial -\overline{\partial })\rho
_{11}|_{S}=-2i\overline{\partial }\rho _{11}|_{S} \\ 
\\ 
n=2i\widetilde{\partial }\rho _{22}|_{S}=i(\widetilde{\partial }-\overline{%
\widetilde{\partial }})\rho _{22}|_{S}=-2i\overline{\widetilde{\partial }}%
\rho _{22}|_{S} \\ 
\\ 
m_{1}=i(\partial +\widetilde{\partial }-\overline{\partial }-\overline{%
\widetilde{\partial }})\frac{\rho _{12}+\overline{\rho _{12}}}{2}|_{S} \\ 
\\ 
m_{2}=i(\partial +\widetilde{\partial }-\overline{\partial }-\overline{%
\widetilde{\partial }})\frac{\overline{\rho _{12}}-\rho _{12}}{2i}|_{S} \\ 
\end{array}
\label{k1}
\end{equation}%
These forms restricted on the manifold are real, because of $d\rho _{ij}=0$
and the special dependence of each function on the structure coordinates $%
\left( z^{\alpha },z^{\widetilde{\alpha }}\right) $. The relations become
simpler if we use the complex form \ 
\begin{equation}
\begin{array}{l}
m=m_{1}+im_{2}=2i\partial \overline{\rho _{12}}=-2i\overline{\widetilde{%
\partial }}\overline{\rho _{12}}=i(\partial -\overline{\widetilde{\partial }}%
)\overline{\rho _{12}} \\ 
\end{array}
\label{k2}
\end{equation}%
Notice that these forms coincide with the null tetrad up to a multiplicative
factor. The tetrad-Weyl transformation is implied by the following permitted
transformation \ 
\begin{equation}
\begin{array}{l}
\rho _{11}^{\prime }=\Lambda \rho _{11}\quad ,\quad \rho _{22}^{\prime
}=N\rho _{22}\quad ,\quad \rho _{12}^{\prime }=\overline{M}\rho _{12} \\ 
\end{array}
\label{k3}
\end{equation}%
Where $\Lambda ,N$ and $M$ are general functions which do not vanish in the
definition neighborhood. The general CR transformation is actually
restricted to a factor, because the dimension of the manifold coincides with
its codimension\cite{BAOU}.

In the interesting generic case, the conditions (\ref{p6}) define a
maximally totally real submanifold $S$ of $%
\mathbb{C}
^{4}$. For every such condition and in the corresponding neighborhood of $%
\mathbb{C}
^{4}$ we may define a kaehlerian metric, which turns out to be lorentzian on
the manifold $S$.

I consider the following Kaehler metric \ 
\begin{equation}
\begin{array}{l}
ds^{2}=\tsum\limits_{a,b}\frac{\partial ^{2}\rho }{\partial z^{a}\partial 
\overline{z^{b}}}dz^{a}d\overline{z^{b}} \\ 
\end{array}
\label{k4}
\end{equation}%
where \ 
\begin{equation}
\begin{array}{l}
\rho =\rho _{11}\rho _{22}-\rho _{12}\overline{\rho _{12}} \\ 
\end{array}
\label{k5}
\end{equation}%
A straightforward calculation gives ($f_{a\overline{b}}=\frac{\partial
^{2}\rho }{\partial z^{a}\partial \overline{z^{b}}}$) \ 
\begin{equation}
\begin{array}{l}
f_{a\overline{b}}=\rho _{22}\frac{\partial ^{2}\rho _{11}}{\partial
z^{a}\partial \overline{z^{b}}}+\frac{\partial \rho _{11}}{\partial z^{a}}%
\frac{\partial \rho _{22}}{\partial \overline{z^{b}}}+\frac{\partial \rho
_{22}}{\partial z^{a}}\frac{\partial \rho _{11}}{\partial \overline{z^{b}}}%
+\rho _{11}\frac{\partial ^{2}\rho _{22}}{\partial z^{a}\partial \overline{%
z^{b}}}- \\ 
\qquad -\overline{\rho _{12}}\frac{\partial ^{2}\rho _{12}}{\partial
z^{a}\partial \overline{z^{b}}}-\frac{\partial \overline{\rho _{12}}}{%
\partial z^{a}}\frac{\partial \rho _{12}}{\partial \overline{z^{b}}}-\frac{%
\partial \rho _{12}}{\partial z^{a}}\frac{\partial \overline{\rho _{12}}}{%
\partial \overline{z^{b}}}-\rho _{12}\frac{\partial ^{2}\overline{\rho _{12}}%
}{\partial z^{a}\partial \overline{z^{b}}} \\ 
\end{array}
\label{k6}
\end{equation}

On the surface ($\rho _{ij}=0$) the metric takes the lorentzian form \ 
\begin{equation}
\begin{array}{l}
ds^{2}|_{S}=2(\frac{\partial \rho _{11}}{\partial z^{a}}\frac{\partial \rho
_{22}}{\partial \overline{z^{b}}}+\frac{\partial \rho _{22}}{\partial z^{a}}%
\frac{\partial \rho _{11}}{\partial \overline{z^{b}}}-\frac{\partial 
\overline{\rho _{12}}}{\partial z^{a}}\frac{\partial \rho _{12}}{\partial 
\overline{z^{b}}}-\frac{\partial \overline{\rho _{12}}}{\partial z^{a}}\frac{%
\partial \rho _{12}}{\partial \overline{z^{b}}})dz^{a}d\overline{z^{b}}= \\ 
\\ 
\qquad =2(\ell \otimes n-m\otimes \overline{m}) \\ 
\end{array}
\label{k7}
\end{equation}%
where $\ell ,n$ and $m$ are defined in (\ref{k1}).

Using the $G_{2,2}$ homogeneous coordinates $X^{ni}$, the general defining
relations take the form%
\begin{equation}
\rho =X^{\dagger }EX-%
\begin{pmatrix}
G_{11} & G_{12} \\ 
\overline{G_{12}} & G_{22}%
\end{pmatrix}%
=0  \label{k8}
\end{equation}%
from which we find\cite{RAG2011} \ 
\begin{equation}
\begin{array}{l}
y^{a}=\frac{1}{2\sqrt{2}}[G_{22}N^{a}+G_{11}L^{a}-G_{12}M^{a}-\overline{%
G_{12}}\overline{M}^{a}] \\ 
\end{array}
\label{k9}
\end{equation}%
where $y^{a}$\ is the imaginary part of $r^{a}=x^{a}+iy^{a}$\ defined by the
relation $r_{A^{\prime }B}=r^{a}\sigma _{aA^{\prime }B}$\ \ and the null
tetrad is 
\begin{equation}
\begin{array}{l}
L^{a}=\frac{1}{\sqrt{2}}\overline{\lambda }^{A^{\prime }1}\lambda
^{B1}\sigma _{A^{\prime }B}^{a}\quad ,\quad N^{a}=\frac{1}{\sqrt{2}}%
\overline{\lambda }^{A^{\prime }2}\lambda ^{B2}\sigma _{A^{\prime
}B}^{a}\quad ,\quad M^{a}=\frac{1}{\sqrt{2}}\overline{\lambda }^{A^{\prime
}2}\lambda ^{B1}\sigma _{A^{\prime }B}^{a} \\ 
\\ 
\epsilon _{AB}\lambda ^{A1}\lambda ^{B2}=1 \\ 
\end{array}
\label{k10}
\end{equation}%
If we substitute the normalized $\lambda ^{Ai}$\ as functions of $r^{a}$,
using the Kerr conditions $K_{i}(X^{mi})$, these relations turn out to be
four real functions of $x^{a}$ and $y^{a}$. The implicit function theorem
assures the existence of a solution $y^{a}=$ $h^{a}(x)$ for (\ref{k9}). It
is clear that in the context of the present model gravity is a manifestation
of the penetration of the surface (spacetime) inside the classical domain
and not of the metric tensor. There are spacetimes with nonvanishing
curvature tensor which are compatible with a flat spacetime complex
structure.

Notice that the structure coordinates $z^{a}$ are related to the $G_{2,2}$
projective coordinates $r^{b}$ with holomorphic transformations. Therefore
in the case of the simple condition $X^{\dagger }EX=0$ we can always choose
a tetrad-Weyl transformation such that the Kaehler metric takes the form \ 
\begin{equation}
\begin{array}{l}
ds^{2}=\frac{1}{2}\tsum\limits_{a,b}\frac{\partial ^{2}(-(\overline{r^{c}}%
-r^{c})^{2})}{\partial r^{a}\partial \overline{r^{b}}}dr^{a}d\overline{r^{b}}%
=\eta _{ab}dr^{a}d\overline{r^{b}} \\ 
\end{array}
\label{k11}
\end{equation}%
which apparently becomes the Minkowski metric on the surface $y^{a}=\func{Im}%
(r^{a})=0$.

Therefore we conclude that the spacetime with two geodetic and shear free
congruences is always a submanifold of a Kaehler manifold. In fact it is a
lagrangian submanifold of the corresponding symplectic manifold. This opens
up a way to apply geometric quantization directly to the surfaces, without
reference to the conventional Dirac or BRST quantization of the model.

The relation (\ref{k9}) implies that $y^{a}y^{b}\eta _{ab}<0$\ for any
asymptotically flat spacetime. But taking into account the regularity of the
surface and its holomorphic translation inside the classical (Siegel) domain
with a complex time translation 
\begin{equation}
\begin{array}{l}
\begin{pmatrix}
\lambda ^{\prime Aj} \\ 
w_{B^{\prime }}^{\prime j}%
\end{pmatrix}%
=%
\begin{pmatrix}
I & 0 \\ 
dI & I%
\end{pmatrix}%
\begin{pmatrix}
\lambda ^{Aj} \\ 
w_{B^{\prime }}^{j}%
\end{pmatrix}
\\ 
\end{array}
\label{k12}
\end{equation}
we may restrict the phase space (the Kaehler manifold) to the $SU(2,2)$
classical domain. This permit us to define the necessary finite measure
state line bundle\cite{WOOD}.

\section{PERSPECTIVES}

\setcounter{equation}{0}

Renormalizability seems to be the cornerstone for the unification of gravity
with the other forces of nature. The Einstein action is not renormalizable,
while actions with higher order derivatives are not unitary. The quite
extended hope that superstring model would describe phenomenology was based
on its consistency with Quantum Theory. On the other hand conventional
Quantum Field Theoretic models, where every elementary particle is
represented by a field with a corresponding quadratic term in the action,
seems to have reached its limitations with the Standard Model. Any attempt
to make it generally covariant introduces the metric which generates
geometric counterterms in the action. My proposal is to skip from the metric
to the complex structure of the spacetime. The present model is an example
of a four dimensional renormalizable model which depends on the lorentzian
complex structure of the spacetime and not on its compatible metrics (\ref%
{i11)}. This step opens up a new branch in four dimensional Quantum Field
Theory, which is analogous to the two dimensional Conformal Field Theories.

The "colorless particles" of the model are special four dimensional open
surfaces inside the $SU(2,2)$ classical domain. The surfaces with at least
two geodetic and shear free congruences (that is an integrable tetrad) which
are periodic (by identifying $\mathfrak{I}^{\mathfrak{+}}$ and $\mathfrak{I}%
^{\mathfrak{-}}$) represent the vacuum sector of the model. These are the 12
variables of the lorentzian complex structure arranged into representations
of the Poincar\'{e} group. I think the most important step of this kind of
complex structure based models would be the explicit (formal) derivation of
the field representations of these modes. The non-periodic solitonic
configurations constitute the "leptonic" sector of the model. The ordinary
Kerr-Newman lorentzian complex structure cannot be an acceptable solitonic
solution, because it is singular at $r=0$. But a simple asymptotic
calculation indicates that its mass and spin measure the non-periodicity of
the tetrad.

Minkowski spacetime coincides with the characteristic (Shilov) boundary of
the classical domain. The spin and the gravity of the surface measures how
much deep inside the classical domain the surface penetrates. The lorentzian
complex structure does not uniquely determine the metric of the surface. But
the metric, which admits a lorentzian complex structure, determines it
through the algebraic condition $\Psi _{0}=\Psi _{4}=0$. I proved that every
surface is a totally real lagrangian submanifold of a Kaehler (symplectic)
ambient manifold. The permitted metrics of the surface are restrictions of
corresponding Kaehler potentials. The consequences of the geometric
quantization of some special surfaces is under investigation. My expectation
is that this quantization will fix the spontaneous breaking of the
tetrad-Weyl symmetry. That is, it will fix the Kaehler potential and
subsequently the spacetime metric.

A static soliton generated by one simple complex trajectory represents a
"particle" and its complex conjugate structure defines its "antiparticle".
The $g=2$ gyromagnetic ratio assures that the solitonic particle is
fermionic. A surface with at least two coordinate neighborhoods and
particle-like asymptotic behavior, represents the scattering of these
"particles". Recall that it is exactly the Einstein point of view that
generated the equations of motion. This implies that the Einstein equations
should be viewed as the definition of the energy. But it is not yet clear
how Quantum Mechanics breaks the tetrad-Weyl symmetry to provide a unique
tetrad and subsequently metric to the surface.

Hence we conclude that the string model is not unique. Therefore the
nonobservance of supersymmetric particles is not a setback to the process of
unifying gravity with the other forces in nature. Besides the present model
is more conventional and it has many more natural features than the string
model.

\bigskip
\begin{verbatim}
 
\end{verbatim}

\end{document}